\documentclass[preprintnumbers,article,amsmath,amssymb,floatfix,10pt,prd,onecolumn,
superscriptaddress,nofootinbib]{revtex4-2}
\usepackage{bm}
\usepackage{amsfonts}
\usepackage{latexsym}
\usepackage[latin1]{inputenc}
\usepackage{graphicx}
\usepackage{amsmath}
\usepackage{palatino}
\usepackage{mathpazo}
\usepackage{textcomp}
\usepackage{float}
\usepackage{booktabs}
\usepackage{dcolumn}
\usepackage{ragged2e}
\usepackage{hyperref}
\hypersetup{colorlinks,citecolor=blue}
\hypersetup{colorlinks=true,linkcolor=red,filecolor=magenta,    urlcolor=blue}
\usepackage{amsmath}
\usepackage{xcolor}
\usepackage{orcidlink}
\usepackage{epsfig}
\usepackage{subfigure}
\usepackage{commath}

\def\jnl@style{\it}
\def\aaref@jnl#1{{\jnl@style#1}}

\def\aaref@jnl#1{{\jnl@style#1}}

\def\aj{\aaref@jnl{AJ}}                   
\def\apj{\aaref@jnl{ApJ}}                 
\def\apjl{\aaref@jnl{ApJ}}                
\def\apjs{\aaref@jnl{ApJS}}               
\def\apss{\aaref@jnl{Ap\&SS}}             
\def\aap{\aaref@jnl{A\&A}}                
\def\aapr{\aaref@jnl{A\&A~Rev.}}          
\def\aaps{\aaref@jnl{A\&AS}}              
\def\mnras{\aaref@jnl{Mon.~Not.~Roy.~Astron.~Soc.}}             
\def\prd{\aaref@jnl{Phys.~Rev.~D}}        
\def\prc{\aaref@jnl{Phys.~Rev.~C}}  
\def\prl{\aaref@jnl{Phys.~Rev.~Lett.}}    
\def\qjras{\aaref@jnl{QJRAS}}             
\def\skytel{\aaref@jnl{S\&T}}             
\def\ssr{\aaref@jnl{Space~Sci.~Rev.}}     
\def\zap{\aaref@jnl{ZAp}}                 
\def\nat{\aaref@jnl{Nature}}              
\def\aplett{\aaref@jnl{Astrophys.~Lett.}} 
\def\apspr{\aaref@jnl{Astrophys.~Space~Phys.~Res.}} 
\def\physrep{\aaref@jnl{Phys.~Rep.}}      
\def\physscr{\aaref@jnl{Phys.~Scr}}       
\def\commat{\aaref@jnl{Comm.~Math.~Phys.}}              
\def\science{\aaref@jnl{Science}}               
\def\cqg{\aaref@jnl{Classical Quant.~Grav.}}            
\def\jpcs{\aaref@jnl{JPCS}}                                     
\def\ijmpd{\aaref@jnl{Int.~J.~Mod.~Phys.~D}}                    
\def\grg{\aaref@jnl{Gen.~Relat.~Gravit.}}               
\def\rpp{\aaref@jnl{Rep.~Prog.~Phys.}}          
\def\npa{\aaref@jnl{Nucl.~Phys.~A}}        
\def\lrr{\aaref@jnl{Living Rev.~Rel.}}                   
\def\jcap{\aaref@jnl{J.~Cosmology Astropart.~Phys.}}    
\def\rmp{\aaref@jnl{Rev.~Mod.~Phys.}}   
\def\epjc{\aaref@jnl{Eur.~Phys.~J.~C}} 
\def\plb{\aaref@jnl{~Phy.~Lett.~B}} 
\def\mpla{\aaref@jnl{Mod.~Phy.~Lett.~A}} 
\def\arxiv{\aaref@jnl{arxiv.org}}

\allowdisplaybreaks[1]

\begin{document}
\color{black}       
\title{Static spherically symmetric wormholes in $f(Q,T)$ gravity}

\author{Moreshwar Tayde\orcidlink{0000-0002-3110-3411}}
\email{moreshwartayde@gmail.com}
\affiliation{Department of Mathematics, Birla Institute of Technology and
Science-Pilani,\\ Hyderabad Campus, Hyderabad-500078, India.}

\author{Zinnat Hassan\orcidlink{0000-0002-6608-2075}}
\email{zinnathassan980@gmail.com}
\affiliation{Department of Mathematics, Birla Institute of Technology and
Science-Pilani,\\ Hyderabad Campus, Hyderabad-500078, India.}

\author{P.K. Sahoo\orcidlink{0000-0003-2130-8832}}
\email{pksahoo@hyderabad.bits-pilani.ac.in}
\affiliation{Department of Mathematics, Birla Institute of Technology and
Science-Pilani,\\ Hyderabad Campus, Hyderabad-500078, India.}

\author{Sashideep Gutti}
\email{sashideep@hyderabad.bits-pilani.ac.in}
\affiliation{Department of Physics, Birla Institute of Technology and
Science-Pilani,\\ Hyderabad Campus, Hyderabad-500078, India.}

%
\date{\today}
\begin{abstract}
In this article we obtain wormhole solutions in the recently proposed extension of symmetric teleparallel gravity called  $f(Q,T)$ gravity. Here, the gravitational Lagrangian $L$ is defined by an arbitrary function $f$ of  $Q$ and $T$ (where $Q$ is the non-metricity scalar, while $T$ is the trace of the energy-momentum tensor). In this study, we obtain the field equations for a  static spherically symmetric wormhole metric in the context of a general $f(Q,T)$ gravity. We study the wormhole solutions with (i) linear EoS and (ii) anisotropy relation. We adopt two different forms of $f(Q,T)$ (a) linear $f(Q,T)=\alpha Q+\beta T$ and (b) non-linear $f(Q,T)=Q+\lambda Q^2+\eta T$ to investigate these solutions. We investigate the various energy conditions to look for preservation and violation among the solutions that we obtained. We find that NEC is violated in both cases of our assumed forms of $f(Q,T)$. Finally, we perform the stability analysis using Tolman-Oppenheimer-Volkov (TOV) equation.
\end{abstract}

\maketitle


\textbf{Keywords:} Wormhole, EoS, energy conditions, stability analysis, $f(Q,T)$ gravity.
\section{Introduction}
Black holes (BHs) and wormholes (WHs) are two fascinating solutions to General Relativity (GR). The evidence of BHs has already been shown in the literature \cite{Abbott,Abbottt,Akiyama} and is established beyond doubt. The discovery of gravity waves and the signatures they carry can soon bring blackholes and it's study to the domain of observational astrophysics.  The presence of wormholes, has till now not yet established, and even their existence is a highly debatable. A detailed study has been done about its existence in \cite{Khatsymovsky}. There is a fundamental difference between both these entities. Blackhole formation is generic and is omnipresent in nature due to the gravitational collapse of stars. There is no special requirement regarding the matter content that facilitates the formation of blackholes. The energy conditions are very much preserved for the formation of blackholes. Wormholes on the other hand are not generic enough and need special ecosystem in terms of the energy conditions for it's formation. The wormhole needs non trivial matter (exotic matter that violates Null energy condition (NEC)) content for it's maintenance and formation \cite{Visser}.\\
After the discovery of blackholes and the possibility of probing interior the blackholes via gravity waves, the work on wormholes also received a boost \cite{Flamm}. The Wormholes and their conception stem from the "Einstein-Rosen bridge \cite{Einstein}, which was initially thought of as only a formal mathematical result. 
In the work \cite{Wheeler}, Wheeler highlighted the theoretical possibility of using the wormhole to form a bridge between two vastly separated region. 
Morris and Thorne \cite{Morris}, in 1988, observed another class of wormhole solutions that maintain the wormhole throat and hence could be traversable. The throat of these wormholes is maintained open due to a specific type of matter that violates the various energy conditions, especially NEC.\\
Such types of matter are known as the exotic matter that is not part of the standard model of particle physics and is studied in the standard model extensions. This type of matter is required in both dynamic \cite{Dehghani,Garattini,Gonzalez,Hansen}, and static \cite{Jamil,Anabalon,Balakin} wormhole scenarios. In general, the classical matter content satisfy all the energy conditions, whereas when one includes the effects of quantum theory for e.g the Cassimir effect, we have situations where the energy conditions are violated. When one talks of quantum gravity too, one can have scenarios where the classical energy conditions can break down leading to a 'repulsive' big bounce.  In-order to quantify the extent of energy violations, the  \cite{Kar}, Visser et al. designed the volume integral quantifier (VIQ) for quantifying the total averaged null energy condition (ANEC). Moreover, Visser \cite{Visser1,Visser2}  proposed a copy-paste technique to  the usminimize the usage of exotic matter, though this technique is applicable to the exotic fluid at the throat of the wormhole. Kuhfittig gave another solution in \cite{Kuhfittig1,Kuhfittig2} that by imposing the condition, $b^{'}(r)\leq 1$ (where $b(r)$ is the shape function) at the wormhole's throat, the region demanding exotic matter can be made arbitrarily small.\\
In recent years, modified theories of gravity have been growing interest among researchers. These theories are the geometrical extension of Einstein's GR, and these are used to describe the early and late time acceleration of the universe. Much work has already been done on astrophysical objects like wormholes in modified theories, and still, research is going on. Born-Infield theory \cite{Richarte,Eiroa,Shaikh}, Rastall theory \cite{Moradpour}, quadratic gravity \cite{Duplessis}, curvature matter coupling \cite{Garcia,Ditta,Garcia1}, Einstein-Cartan gravity \cite{Bronnikov,Bronnikov1,Mehdizadeh}, and braneworld \cite{Camera,Lobo2,Bronnikov2,Parsaei1,Kar1} are few examples.\\
Lobo and Oliveira \cite{Miguel} have investigated wormhole geometries in the context of $f(R)$ gravity (where $R$ is the Ricci scalar). They used specific shape functions and various equations of state to find the exact wormhole solutions and studied the behavior with energy conditions. In \cite{Azizi}, Azizi studied wormhole geometries in $f(R,T)$ gravity and showed the effective stress-energy that is responsible for violation of the NEC. Also, an interesting study on teleparallel gravity have been done by Bahamonde et al. in Ref. \cite{1}. Moreover, one could check more articles on wormhole geometries in different modified theories of gravity, such as in $f(R,T)$ gravity \cite{P.K. Sahoo,Yousaf/2017,Elizalde,Iqra}, $f(T)$ gravity \cite{Jamil/2013,Shamaila,Harko/2012}, $f(T,T_G)$ gravity \cite{Sharif/2018,Sharif/2017} and so on. A nice model outlining solutions of humanly traversable wormholes when generalizing  physics beyond the standard model of particle physics using Randall Sundram model is given by Maldacena \cite{maldacena} \\
It can be noted that the role of the Equation of state (EoS) in cosmology is significant because it portrays the physical fluid, which is vital to sustaining a geometry. Fluid with a linear EoS and positive energy density is an excellent candidate to describe the expansion of the universe. Also, It was considered that phantom energy is a possible choice to describe the accelerated expansion of the universe, and it is referred to as $\omega=p/\rho<-1$. Caldwell first presented this concept in \cite{Caldwell}. Since phantom energy violates the NEC, which is necessary for a traversable wormhole, hence, this energy may play a vital role in forming wormholes like spacetime. In \cite{Lobo5,Gonzalez 1,Zaslavskii,Lukmanova}, the authors have investigated the physical properties of wormholes by considering the phantom energy into account in the context of GR.\\
Recently, Jimenez et al. \cite{Jimenez} introduced a new class of modified gravity named symmetric teleparallel gravity or $f(Q)$ gravity, where $Q$ is the non-metricity scalar. In this theory, both torsion and curvature will disappear, and hence the gravity only depends on the non-metricity. The affine connection plays a significant role in the symmetric teleparallel gravity rather than physical manifold \cite{Jimenez}. It is also noted that $f(Q)$ gravity is attributed to second-order field equations while in $f(R)$ gravity has fourth-order field equations \cite{Faraoni}. Hence, $f(Q)$ gravity gives an alternate geometric portrayal of gravity, which is regardless equivalent to GR. Readers may check some Refs. on cosmological \cite{Koivisto,Frusciante,Solanki} and astrophysical \cite{Shaun,Lin,Hassan,Wang,Mustafa} objects where the authors have studied deeply in this symmetric teleparallel gravity.\\
Yixin et al. \cite{Y.Xu et al} presented the extension of $f(Q)$ gravity named $f(Q,T)$ gravity which is based on the coupling of non-metricity $Q$ and trace of energy-momentum tensor $T$. In this theory, the gravitational effect connects through non-metricity function $Q$ and manifestations from the quantum field due to the trace of the energy-momentum tensor $T$. Since it was recently proposed, a lot of work has already been done on this gravity based on the theoretical \cite{Najera,Bhattacharjee} and observational \cite{Arora1} aspects. Harko et al. have studied the novel couplings between non-metricity and matter in \cite{T.Harko} and also discussed coupling matter in modified $Q$ gravity in \cite{Harko}. A. Delhom \cite{Delhom} investigated minimal coupling in the presence of torsion and non-metricity. Also, $f(R)$ gravity, torsion, and non-metricity were studied by Sotiriou in Ref. \cite{Sotiriou}.\\
Motivated by the above studies, we intended to study wormhole solutions in $f(Q,T)$ gravity for spherically symmetric and static configuration.  We further extend our analysis by considering (i) linear EoS relation and (ii) a relation between radial and tangential pressures under the anisotropy case and try to find the exact wormhole solutions for both cases under linear and non-linear models.\\
The overview of the article is organized as follows: In Sec. \ref{sec2}, we have shown the basic formalism of $f(Q,T)$ gravity and the corresponding field equations for wormhole in $f(Q,T)$ are given in Sec. \ref{sec3}. By considering different form of $f(Q,T)$ models, we studied wormhole solutions in Sec. \ref{sec4}. The stability analysis of obtained wormhole solutions using the TOV equation has been discussed in Sec. \ref{sec5} followed by final remarks in Sec. \ref{sec6}.
\\
\section{Basic Field Equations in $f(Q,T)$ gravity}
\label{sec2}
We consider the action for symmetric teleparallel gravity proposed in \cite{Y.Xu et al}
\begin{equation}\label{1}
\mathcal{S}=\int\frac{1}{16\pi}\,f(Q,T)\sqrt{-g}\,d^4x+\int \mathcal{L}_m\,\sqrt{-g}\,d^4x\, ,
\end{equation}
where $f(Q,T)$ is a function of non-metricity $Q$ and trace of the energy momentum tensor $T$, $g$ is the determinant of the metric $g_{\mu\nu}$, and $\mathcal{L}_m$ is the matter Lagrangian density.\\
The non-metricity tensor is given by \cite{Jimenez}\\
\begin{equation}\label{2}
Q_{\lambda\mu\nu}=\bigtriangledown_{\lambda} g_{\mu\nu},
\end{equation}
Also, we can define the non-metricity conjugate or superpotential as
\begin{equation}\label{4}
P^\alpha\;_{\mu\nu}=\frac{1}{4}\left[-Q^\alpha\;_{\mu\nu}+2Q_{(\mu}\;^\alpha\;_{\nu)}+Q^\alpha g_{\mu\nu}-\tilde{Q}^\alpha g_{\mu\nu}-\delta^\alpha_{(\mu}Q_{\nu)}\right],
\end{equation}
where
\begin{equation}
\label{3}
Q_{\alpha}=Q_{\alpha}\;^{\mu}\;_{\mu},\; \tilde{Q}_\alpha=Q^\mu\;_{\alpha\mu}.
\end{equation}
are two traces of the non-metricity tensor.\\
The non-metricity scalar represented as \cite{Jimenez}
\begin{eqnarray}
\label{5}
Q &=& -Q_{\alpha\mu\nu}\,P^{\alpha\mu\nu}\\
&=& -g^{\mu\nu}\left(L^\beta_{\,\,\,\alpha\mu}\,L^\alpha_{\,\,\,\nu\beta}-L^\beta_{\,\,\,\alpha\beta}\,L^\alpha_{\,\,\,\mu\nu}\right),
\end{eqnarray}
where the disformation $L^\beta_{\,\,\,\mu\nu}$ is defined by
\begin{equation}
L^\beta_{\,\,\,\mu\nu}=\frac{1}{2}Q^\beta_{\,\,\,\mu\nu}-Q_{(\mu\,\,\,\,\,\,\nu)}^{\,\,\,\,\,\,\beta}.
\end{equation}

Now, by varying the action with respect to the metric tensor $g_{\mu\nu}$ one can arrive at the gravitational equations of motion, which can be written as
\begin{equation}\label{7}
\frac{-2}{\sqrt{-g}}\bigtriangledown_\alpha\left(\sqrt{-g}\,f_Q\,P^\alpha\;_{\mu\nu}\right)-\frac{1}{2}g_{\mu\nu}f \\
+f_T \left(T_{\mu\nu} +\Theta_{\mu\nu}\right)\\
-f_Q\left(P_{\mu\alpha\beta}\,Q_\nu\;^{\alpha\beta}-2\,Q^
{\alpha\beta}\,\,_{\mu}\,P_{\alpha\beta\nu}\right)=8\pi T_{\mu\nu},
\end{equation}

where $f_Q=\frac{df}{dQ}$ and $f_T=\frac{df}{dT}$.

By definition, the energy-momentum tensor for the fluid depiction of the spacetime can be composed as
\begin{equation}\label{6}
T_{\mu\nu}=-\frac{2}{\sqrt{-g}}\frac{\delta\left(\sqrt{-g}\,\mathcal{L}_m\right)}{\delta g^{\mu\nu}},
\end{equation}
and
\begin{equation}\label{6a}
\Theta_{\mu\nu}=g^{\alpha\beta}\frac{\delta T_{\alpha\beta}}{\delta g^{\mu\nu}}.
\end{equation}
%
\section{Wormhole in $f(Q,T)$ gravity}
\label{sec3}
Consider the spherically symmetric and static wormhole metric in Schwarzschild coordinates $(t,\,r,\,\theta,\,\Phi)$ is given by \cite{Morris,Visser}
\begin{equation}\label{8}
ds^2=e^{2\phi(r)}dt^2-\left(1-\frac{b(r)}{r}\right)^{-1}dr^2-r^2\,d\theta^2-r^2\,\sin^2\theta\,d\Phi^2,
\end{equation}
where $\phi(r)$ and $b(r)$ denotes the redshift function and the shape function, respectively. Also, both obey the following conditions \cite{Morris,Visser}:
\begin{itemize}
  \item[(1)] For $r>r_0$, i.e., out of throat, $1-\frac{b(r)}{r}>0$, and at the wormhole's throat i.e., $r=r_0$, $b(r)$ must satisfy the condition $b(r_0)=r_0$.
  \item[(2)] The shape function $b(r)$ has to fulfill the flaring-out requirement at the throat i.e., $b'(r_0)<1$.
  \item[(3)] For asymptotic flatness condition, the limit $\frac{b(r)}{r}\rightarrow 0$ as $r\rightarrow \infty$ is required.
  \item[(4)] Redshift function $\phi(r)$ should be finite everywhere.
\end{itemize}
In this coordinate system, the nonvanishing components of $Q_{\lambda\mu\nu}$ and $L^{\lambda}_{\,\,\,\mu\nu}$ are
\begin{equation}
Q_{rtt}=2\,e^{2\phi}\phi^{'},\,\,\,\,\,Q_{rrr}=-\frac{(rb^{'}-b)}{(r-b)^2},\,\,\,\,\,Q_{\theta r \theta}=Q_{\theta\theta r}=-\frac{rb}{r-b},\,\,\,\,Q_{\Phi r \Phi}=Q_{\Phi\Phi\,r}=-\frac{rb\sin^2\theta}{r-b},
\end{equation}
and
\begin{equation}
L^{t}_{\,\,t\,r}=L^{t}_{\,\,r\,t}=-\phi^{'},\,\,\,\,L^{r}_{\,\,\theta\theta}=-b,\,\,\,L^{r}_{\,\,t\,t}=-\frac{(r-b)}{r}e^{2\phi}\phi{'},\,\,\,L^{r}_{\,\,r\,r}=-\frac{(rb^{'}-b}{2\,r(r-b)},\,\,L^{r}_{\,\,\Phi\Phi}=-b\sin^2\theta,
\end{equation}
respectively, where $'$ represents derivative with respect to the radial coordinate $r$.\\
In the present study, we assume the matter content described by an anisotropic energy-momentum tensor to analyze the wormhole solutions which is given by \cite{Morris,Visser}
\begin{equation}\label{10}
T_{\mu}^{\nu}=\left(\rho+p_t\right)u_{\mu}\,u^{\nu}-p_t\,\delta_{\mu}^{\nu}+\left(p_r-p_t\right)v_{\mu}\,v^{\nu},
\end{equation}
where, $\rho$ denotes the energy density. $u_{\mu}$ and $v_{\mu}$ are the four velocity vector and unitary space-like vectors, respectively. Also both are satisfy the conditions $u_{\mu}u^{\nu}=-v_{\mu}v^{\nu}=1$. $p_r$ and $p_t$ denotes the radial and tangential pressures and both are function of radial coordinate $r$. The trace of the energy-momentum tensor turns out to be $T=\rho-p_r-2p_t$.\\
In this article, we consider matter  Lagrangian $\mathcal{L}_m=-P$ \cite{Correa} and hence Eq. \eqref{6a} can be reads
\begin{equation}
    \Theta_{\mu\nu}=-g_{\mu\nu}\,P-2\,T_{\mu\nu},
\end{equation}
where $P$ is the total pressure and it can be written as $P=\frac{p_r+2\,p_t}{3}$.\\
The non-metricity scalar $Q$ for the metric \eqref{8} is given by
\begin{equation}\label{9}
Q=-\frac{b}{r^2}\left[\frac{rb^{'}-b}{r(r-b)}+2\phi^{'}\right].
\end{equation}
\\
Now, inserting the Eqs. \eqref{8}, \eqref{10} and \eqref{9} into the motion equation \eqref{7}, we get the non-zero components of the field equations for $f(Q,T)$ gravity are given by
\begin{equation}\label{11}
\frac{2 (r-b)}{(2 r-b) f_Q}\left[\rho-\frac{(r-b)}{8 \pi  r^3} \left(\frac{b r f_{\text{QQ}} Q'}{r-b}+b f_Q \left(\frac{ r \phi '+1}{r-b}-\frac{2 r-b}{2 (r-b)^2}\right)+\frac{f r^3}{2 (r-b)}\right)+\frac{f_T (P+\rho )}{8 \pi }\right]=\frac{b'}{8 \pi  r^2},
\end{equation}
\begin{multline}\label{12}
\frac{2 b}{f r^3}\left[p_r +\frac{(r-b)}{16 \pi r^3} \left(f_Q \left(\frac{b \left(\frac{r b'-b}{r-b}+2 r \phi '+2\right)}{r-b}-4 r \phi '\right)+\frac{2 b r f_{\text{QQ}} Q'}{r-b}\right)+\frac{fr^3 (r-b)\phi '}{8\pi b  r^2}-\frac{f_T \left(P-p_r\right)}{8 \pi }\right]\\
=\frac{1}{8 \pi}\left[2\left(1-\frac{b}{r}\right)\frac{\phi '}{r}-\frac{b}{r^3}\right],
\end{multline}
\begin{multline}\label{13}
\frac{1}{f_Q \left(\frac{r}{r-b}+r \phi '\right)}\left[p_t +\frac{(r-b)}{32 \pi r^2}\left(f_Q \left(\frac{4 (2 b-r) \phi '}{r-b}-4 r \left(\phi '\right)^2-4 r \phi ''\right)+\frac{2 f r^2}{r-b}-4 r f_{\text{QQ}} Q' \phi '\right) \right.\\ \left.
+\frac{(r-b)}{8\pi r}\left(\phi '' +{\phi '}^2-\frac{(rb'-b)\phi '}{2r(r-b)}+\frac{\phi '}{r}\right)f_Q \left(\frac{r}{r-b}+r \phi '\right)-\frac{f_T \left(P-p_t\right)}{8 \pi }\right]\\
=\frac{1}{8\pi}\left(1-\frac{b}{r}\right)\left[\phi '' +{\phi '}^2-\frac{(rb'-b)\phi '}{2r(r-b)}-\frac{rb'-b}{2r^2 (r-b)}+\frac{\phi '}{r}\right].
\end{multline}\\
It is known that the Morris-Throne traversable wormhole's field equations for GR can be written as
\begin{equation}\label{13a}
\frac{b'}{8 \pi  r^2}= \tilde{\rho},
\end{equation}
\begin{equation}\label{13b}
\frac{1}{8 \pi}\left[2\left(1-\frac{b}{r}\right)\frac{\phi '}{r}-\frac{b}{r^3}\right] = \tilde{p_r},
\end{equation}
\begin{equation}\label{13c}
\frac{1}{8\pi}\left(1-\frac{b}{r}\right)\left[\phi '' +{\phi '}^2-\frac{(rb'-b)\phi '}{2r(r-b)}-\frac{rb'-b}{2r^2 (r-b)}+\frac{\phi '}{r}\right] = \tilde{p_t}.
\end{equation}
where $\tilde{\rho}$,$\tilde{p_r}$ and $\tilde{p_t}$ are corresponding energy density, radial pressure and tangential pressure respectively and comparing Eqs. \eqref{11}-\eqref{13} with the Eqs. \eqref{13a}-\eqref{13c}, we get
\begin{equation}\label{14}
\tilde{\rho}=\frac{2 (r-b)}{(2 r-b) f_Q}\left[\rho-\frac{1}{8 \pi  r^2}\left(1-\frac{b}{r}\right) \left(\frac{b r f_{\text{QQ}} Q'}{r-b}+b f_Q \left(\frac{ r \phi '+1}{r-b}-\frac{2 r-b}{2 (r-b)^2}\right)+\frac{f r^3}{2 (r-b)}\right)+\frac{f_T (P+\rho )}{8 \pi }\right],
\end{equation}
\begin{equation}\label{15}
\tilde{p_r}=\frac{2 b}{f r^3}\left[p_r +\frac{1}{16 \pi r^2}\left(1-\frac{b}{r}\right) \left(f_Q \left(\frac{b \left(\frac{r b'-b}{r-b}+2 r \phi '+2\right)}{r-b}-4 r \phi '\right)+\frac{2 b r f_{\text{QQ}} Q'}{r-b}\right)+\frac{fr^3 (r-b)\phi '}{8\pi b  r^2}-\frac{f_T \left(P-p_r\right)}{8 \pi }\right],
\end{equation}
\begin{multline}\label{16}
\tilde{p_t}=\frac{1}{f_Q \left(\frac{r}{r-b}+r \phi '\right)}\left[p_t +\frac{1}{32 \pi r}\left(1-\frac{b}{r}\right) \left(f_Q \left(\frac{4 (2 b-r) \phi '}{r-b}-4 r \left(\phi '\right)^2-4 r \phi ''\right)+\frac{2 f r^2}{r-b}-4 r f_{\text{QQ}} Q' \phi '\right) \right.\\ \left.
+\frac{1}{8\pi}\left(1-\frac{b}{r}\right)\left(\phi '' +{\phi '}^2-\frac{(rb'-b)\phi '}{2r(r-b)}+\frac{\phi '}{r}\right)f_Q \left(\frac{r}{r-b}+r \phi '\right)-\frac{f_T \left(P-p_t\right)}{8 \pi }\right].
\end{multline}\\
From the above equations we can see the extra terms beyond GR equations. Also, considering Eqs. \eqref{11}-\eqref{13}, one can find the corresponding field equations for $f(Q,T)$ gravity are given below
\begin{equation}\label{17}
8 \pi  \rho =\frac{(r-b)}{2 r^3} \left[f_Q \left(\frac{(2 r-b) \left(r b'-b\right)}{(r-b)^2}+\frac{b \left(2 r \phi '+2\right)}{r-b}\right)+\frac{2 b r f_{\text{QQ}} Q'}{r-b}+\frac{f r^3}{r-b}-\frac{2r^3 f_T (P+\rho )}{(r-b)}\right],
\end{equation}
\begin{equation}\label{18}
8 \pi  p_r=-\frac{(r-b)}{2 r^3} \left[f_Q \left(\frac{b }{r-b}\left(\frac{r b'-b}{r-b}+2 r \phi '+2\right)-4 r \phi '\right)+\frac{2 b r f_{\text{QQ}} Q'}{r-b}+\frac{f r^3}{r-b}-\frac{2r^3 f_T \left(P-p_r\right)}{(r-b)}\right],
\end{equation}
\begin{equation}\label{19}
8 \pi  p_t=-\frac{(r-b)}{4 r^2} \left[f_Q \left(\frac{\left(r b'-b\right) \left(\frac{2 r}{r-b}+2 r \phi '\right)}{r (r-b)}+\frac{4 (2 b-r) \phi '}{r-b}-4 r \left(\phi '\right)^2-4 r \phi ''\right)-4 r f_{\text{QQ}} Q' \phi '+\frac{2 f r^2}{r-b}-\frac{4r^2 f_T \left(P-p_t\right)}{(r-b)}\right].
\end{equation}\\
With these field equations, one can study wormhole solutions by considering different models in $f(Q,T)$ gravity.\\

Let us dedicate a few lines to classical energy conditions developed from the Raychaudhuri equations. These conditions are used to discuss the physically realistic matter configuration. The four energy conditions the null energy condition (NEC), weak energy condition (WEC), dominant energy condition (DEC),
and strong energy condition (SEC), are expressed as:\\
$\bullet$ Weak energy conditions (WEC) if $\rho\geq0$, $\rho+p_j\geq0$, $\forall j$.\\
$\bullet$ Null energy condition (NEC) if $\rho+p_j\geq0$, $\forall j$.\\
$\bullet$ Dominant energy conditions (DEC) if $\rho\geq0$, $\rho \pm p_j\geq0$, $\forall j$.\\
$\bullet$ Strong energy conditions (SEC) if $\rho+p_j\geq0$, $\rho+\sum_jp_j\geq0$, $\forall j$,\\
where $j=r,\,t$.\\
It is known that, in GR, The NEC is significant because the violation of the NEC may confirm that there must be present exotic matter in WH's throat. Hence NEC is usually studied for wormhole solutions in GR. Moreover, energy density also needs to be positive for a realistic matter source that maintains the wormhole solutions.\\
Solving equations \eqref{17} to \eqref{19} , one can reads
\begin{multline}\label{20}
\rho =-\frac{f_Q f_T \left(-r b' \left(2 r (r-b) \phi '+b+2 r\right)+3 b^2+4 r (b-r) \left(\phi ' \left(r (b-r) \phi '+3 b-2 r\right)+r (b-r) \phi ''\right)\right)}{48 \pi  r^3 (r-b) \left(f_T+8 \pi \right)}\\
-\frac{24 \pi  f_Q \left(r (b-2 r) b'+b \left(2 r (b-r) \phi '+b\right)\right)}{48 \pi  r^3 (r-b) \left(f_T+8 \pi \right)}\\
-\frac{r (b-r) \left(f_T \left(2 f_{\text{QQ}} Q' \left(2 r (b-r) \phi '+b\right)+3 f r^2\right)+24 \pi  \left(2 b f_{\text{QQ}} Q'+f r^2\right)\right)}{48 \pi  r^3 (r-b) \left(f_T+8 \pi \right)},
\end{multline}
\begin{multline}\label{21}
p_r=\frac{f_Q f_T \left(r b' \left(2 r (r-b) \phi '+b+2 r\right)-3 b^2-4 r (b-r) \left(\phi ' \left(r (b-r) \phi '+3 b-2 r\right)+r (b-r) \phi ''\right)\right)}{48 \pi  r^3 (b-r) \left(f_T+8 \pi \right)}\\
+\frac{24 \pi  f_Q \left(b r b'-(3 b-2 r) \left(2 r (b-r) \phi '+b\right)\right)}{48 \pi  r^3 (b-r) \left(f_T+8 \pi \right)}\\
-\frac{r (b-r) \left(f_T \left(2 f_{\text{QQ}} Q' \left(2 r (b-r) \phi '+b\right)+3 f r^2\right)+24 \pi  \left(2 b f_{\text{QQ}} Q'+f r^2\right)\right)}{48 \pi  r^3 (b-r) \left(f_T+8 \pi \right)},
\end{multline}
\begin{multline}\label{22}
p_t=-\frac{f_Q f_T \left(-r b' \left(2 r (r-b) \phi '+b+2 r\right)+3 b^2+4 r (b-r) \left(\phi ' \left(r (b-r) \phi '+3 b-2 r\right)+r (b-r) \phi ''\right)\right)}{48 \pi  r^3 (b-r) \left(f_T+8 \pi \right)}\\
-\frac{24 \pi  r f_Q \left(r \left(b' \left((b-r) \phi '-1\right)+2 (b-r)^2 \left(\left(\phi '\right)^2+\phi ''\right)+(2 r-5 b) \phi '\right)+b \left(3 b \phi '+1\right)\right)}{48 \pi  r^3 (b-r) \left(f_T+8 \pi \right)}\\
-\frac{r (b-r) \left(f_T \left(2 b f_{\text{QQ}} Q'+3 f r^2\right)+4 r (b-r) f_{\text{QQ}} \left(f_T+12 \pi \right) Q' \phi '+24 \pi  f r^2\right)}{48 \pi  r^3 (b-r) \left(f_T+8 \pi \right)}.
\end{multline}\\
 Therefore, the previous equations establish the
following constraints for the energy conditions:\\\\
$\bullet$ \textbf{WEC :} \begin{multline}
\rho = -\frac{f_Q f_T \left(-r b' \left(2 r (r-b) \phi '+b+2 r\right)+3 b^2+4 r (b-r) \left(\phi ' \left(r (b-r) \phi '+3 b-2 r\right)+r (b-r) \phi ''\right)\right)}{48 \pi  r^3 (r-b) \left(f_T+8 \pi \right)}\\
-\frac{24 \pi  f_Q \left(r (b-2 r) b'+b \left(2 r (b-r) \phi '+b\right)\right)}{48 \pi  r^3 (r-b) \left(f_T+8 \pi \right)}\\
-\frac{r (b-r) \left(f_T \left(2 f_{\text{QQ}} Q' \left(2 r (b-r) \phi '+b\right)+3 f r^2\right)+24 \pi  \left(2 b f_{\text{QQ}} Q'+f r^2\right)\right)}{48 \pi  r^3 (r-b) \left(f_T+8 \pi \right)}\geq0,\\
\rho + p_r = \frac{f_Q \left(r b'+2 r (r-b) \phi '-b\right)}{r^3 \left(f_T+8 \pi \right)}\geq0 ,\\
\rho + p_t = \frac{f_Q \left(r \left(b' \left(1-r \phi '\right)+2 r (r-b) \left(\left(\phi '\right)^2+\phi ''\right)-\left((b-2 r) \phi '\right)\right)+b\right)+2 r f_{\text{QQ}} Q' \left(r (r-b) \phi '+b\right)}{2 r^3 \left(f_T+8 \pi \right)}\geq0.
\end{multline}\\
$\bullet$ \textbf{NEC :} 
\begin{multline}
\rho + p_r = \frac{f_Q \left(r b'+2 r (r-b) \phi '-b\right)}{r^3 \left(f_T+8 \pi \right)}\geq0 ,\\
\rho + p_t = \frac{f_Q \left(r \left(b' \left(1-r \phi '\right)+2 r (r-b) \left(\left(\phi '\right)^2+\phi ''\right)-\left((b-2 r) \phi '\right)\right)+b\right)+2 r f_{\text{QQ}} Q' \left(r (r-b) \phi '+b\right)}{2 r^3 \left(f_T+8 \pi \right)}\geq0.
\end{multline}\\
$\bullet$ \textbf{DEC :} 
\begin{multline}
\rho = -\frac{f_Q f_T \left(-r b' \left(2 r (r-b) \phi '+b+2 r\right)+3 b^2+4 r (b-r) \left(\phi ' \left(r (b-r) \phi '+3 b-2 r\right)+r (b-r) \phi ''\right)\right)}{48 \pi  r^3 (r-b) \left(f_T+8 \pi \right)}\\
-\frac{24 \pi  f_Q \left(r (b-2 r) b'+b \left(2 r (b-r) \phi '+b\right)\right)}{48 \pi  r^3 (r-b) \left(f_T+8 \pi \right)}\\
-\frac{r (b-r) \left(f_T \left(2 f_{\text{QQ}} Q' \left(2 r (b-r) \phi '+b\right)+3 f r^2\right)+24 \pi  \left(2 b f_{\text{QQ}} Q'+f r^2\right)\right)}{48 \pi  r^3 (r-b) \left(f_T+8 \pi \right)}\geq0,\\
\rho - p_r = \frac{f_Q f_T \left(r b' \left(2 r (r-b) \phi '+b+2 r\right)-3 b^2-4 r (b-r) \left(\phi ' \left(r (b-r) \phi '+3 b-2 r\right)+r (b-r) \phi ''\right)\right)}{24 \pi  r^3 (r-b) \left(f_T+8 \pi \right)}\\
+\frac{24 \pi  f_Q \left(r^2 b'-(2 b-r) \left(2 r (b-r) \phi '+b\right)\right)}{24 \pi  r^3 (r-b) \left(f_T+8 \pi \right)}\\
-\frac{r (b-r) \left(f_T \left(2 f_{\text{QQ}} Q' \left(2 r (b-r) \phi '+b\right)+3 f r^2\right)+24 \pi  \left(2 b f_{\text{QQ}} Q'+f r^2\right)\right)}{24 \pi  r^3 (r-b) \left(f_T+8 \pi \right)}\geq0 ,\\
\rho - p_t = -\frac{f_Q f_T \left(-r b' \left(2 r (r-b) \phi '+b+2 r\right)+3 b^2+4 r (b-r) \left(\phi ' \left(r (b-r) \phi '+3 b-2 r\right)+r (b-r) \phi ''\right)\right)}{24 \pi  r^3 (r-b) \left(f_T+8 \pi \right)}\\
-\frac{12 \pi  f_Q \left(r b' \left(r (b-r) \phi '+b-3 r\right)+r (b-r) \left(\phi ' \left(2 r (b-r) \phi '+5 b-2 r\right)+2 r (b-r) \phi ''\right)+b (b+r)\right)}{24 \pi  r^3 (r-b) \left(f_T+8 \pi \right)}\\
-\frac{r (b-r) \left(f_T \left(2 f_{\text{QQ}} Q' \left(2 r (b-r) \phi '+b\right)+3 f r^2\right)+24 \pi  \left(f_{\text{QQ}} Q' \left(r (b-r) \phi '+b\right)+f r^2\right)\right)}{24 \pi  r^3 (r-b) \left(f_T+8 \pi \right)}\geq0.
\end{multline}\\
$\bullet$ \textbf{SEC :} 
\begin{multline}
\rho + p_r + 2 p_t = \frac{f_Q f_T \left(-r b' \left(2 r (r-b) \phi '+b+2 r\right)+3 b^2+4 r (b-r) \left(\phi ' \left(r (b-r) \phi '+3 b-2 r\right)+r (b-r) \phi ''\right)\right)}{24 \pi  r^3 (r-b) \left(f_T+8 \pi \right)}\\
+\frac{24 \pi  f_Q \left(-r b' \left(r (r-b) \phi '+b\right)+b^2+r (b-r) \left(\phi ' \left(2 r (b-r) \phi '+5 b-4 r\right)+2 r (b-r) \phi ''\right)\right)}{24 \pi  r^3 (r-b) \left(f_T+8 \pi \right)}\\
+\frac{r (b-r) \left(f_T \left(2 b f_{\text{QQ}} Q'+3 f r^2\right)+4 r (b-r) f_{\text{QQ}} \left(f_T+12 \pi \right) Q' \phi '+24 \pi  f r^2\right)}{24 \pi  r^3 (r-b) \left(f_T+8 \pi \right)}\geq0.
\end{multline}\\

\section{Wormhole solutions with different $f(Q,T)$ models}\label{sec4}
In this section, we are going to study the wormhole solutions with different functional forms $f(Q,T)$ gravity. We will analyze the behaviors of the solutions with energy conditions. Also, to achieve the de Sitter and anti-de Sitter asymptotic behavior, one could consider redshift function $\phi(r)=constant$. In our whole study, we consider constant redshift function.
\subsubsection{\textbf{Wormhole solutions with linear EoS}}
\begin{figure}[H]
\centering
\subfigure[]{\includegraphics[width=6.5cm,height=4cm]{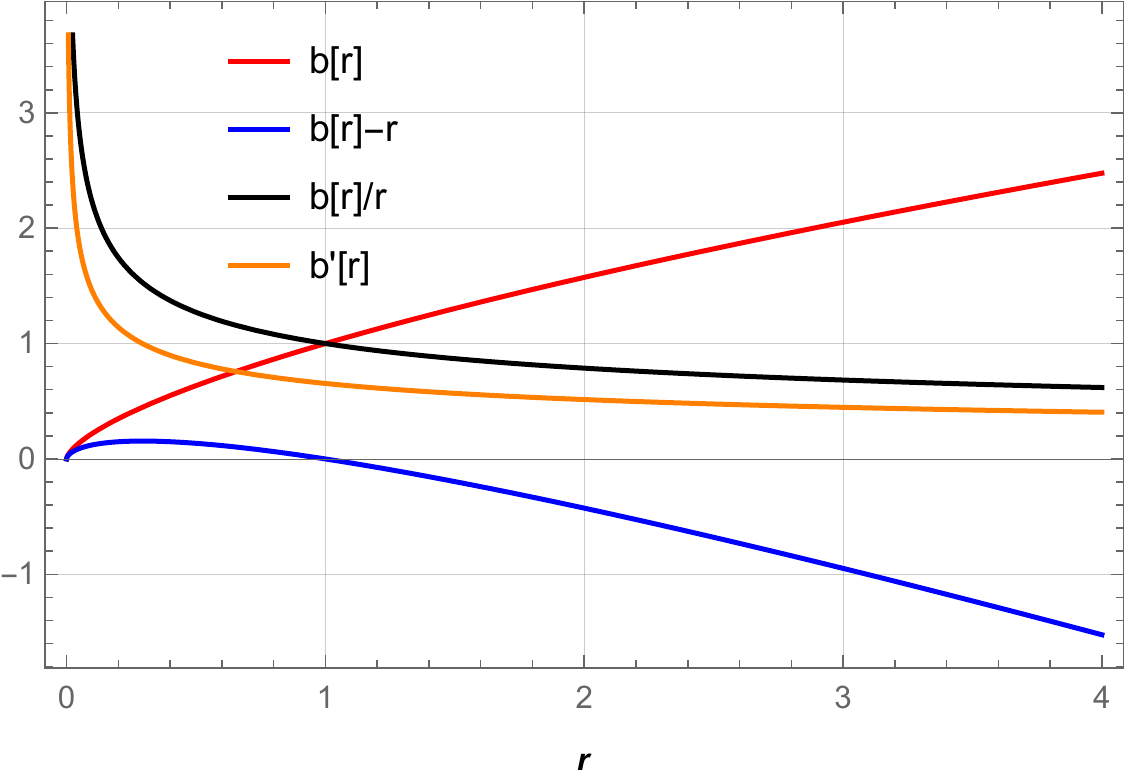}}\,\,\,\,\,\,\,\,\,\,\,\,\,\,\,\,\,\,\,\,\,\,\,\,\,
\subfigure[]{\includegraphics[width=6.5cm,height=4cm]{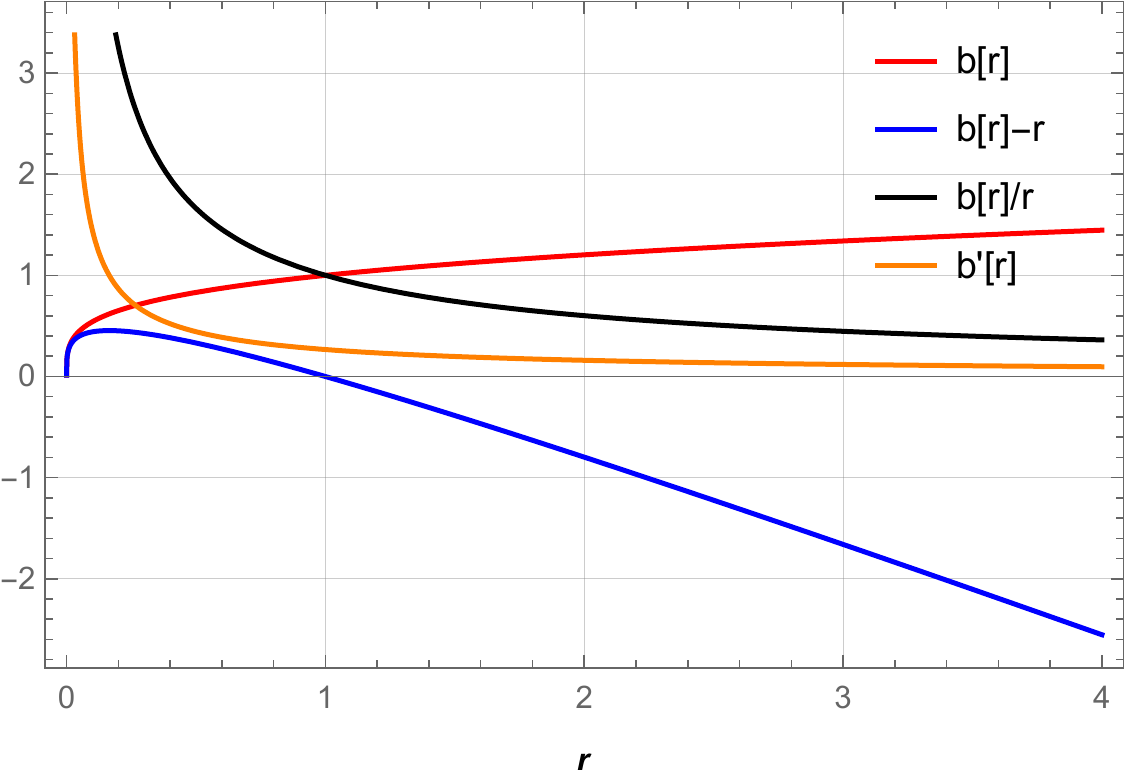}}
\caption{The behavior of shape function $b(r)$, flaring out condition $b'(r)<1$, throat condition $b(r)-r<0$ and asymptotically flatness condition $\frac{b(r)}{r}\rightarrow 0$ as $r\rightarrow \infty$ for (a) $\omega= -1.5$ with $\beta=1$ and (b) $\omega= -0.5$ with $\beta=14$. Also, we consider $\alpha=1$ and used the unit of radius in kilometers (km).}
\label{Fig1}
\end{figure}

In this subsection, we consider EoS (which represents a link between the components of energy-momentum tensor) to solve the field equations and construct the WH solutions. Usually, this concept is used in GR to find the exact WH solutions. Due to the complexity of field equations in the modified gravity, sometimes researchers fail to obtain exact analytical solutions, and hence they use numerical methods, or they do not consider EoS. In literature, a linear EoS $p_r=\omega\rho$ is the most usual EoS for examining the wormhole solutions. In this work, we will study WH solutions with the following form of EoS \cite{Lobo5,Jusufi1}.\\
\begin{equation}
\label{2a}
p_r=\omega\rho,
\end{equation}
where $\omega$ is the EoS parameter. In Ref. \cite{Foad}, the authors mentioned that asymptotically flat WH solutions with $\omega \leq -1$ dubbed as phantom region EoS. Also, it is mentioned in \cite{Hassan} that obtaining WH solutions with linear EoS not easy in symmetric teleparallel gravity. However, in this study, we will try to obtain exact WH solutions in $f(Q,T)$ gravity.\\
By using Eqs. \eqref{20}, \eqref{21} in Eq. \eqref{2a}, we get
\begin{multline}
\label{2b}
f_Q \left[f_T \left(r (b+2 r) b'-3 b^2\right)+24 \pi  \left(b r b'-b (3 b-2 r)\right)\right]-r (b-r) \left[f_T \left(2 b f_{\text{QQ}} Q'+3 f r^2\right)+24 \pi  \left(2 b f_{\text{QQ}} Q'+f r^2\right)\right]=\\
\omega  \left[f_Q \left(f_T \left(3 b^2-r (b+2 r) b'\right)+24 \pi  \left(r (b-2 r) b'+b^2\right)\right)+r (b-r) \left(f_T \left(2 b f_{\text{QQ}} Q'+3 f r^2\right)+24 \pi  \left(2 b f_{\text{QQ}} Q'+f r^2\right)\right)\right].
\end{multline}
Solving the above equation for general form of $f(Q,T)$ is not easy hence we will consider some specific functional forms of $f(Q,T)$ to find the shape function $b(r)$. In this study, we have considered two specific form of $f(Q,T)$ such as linear ($f(Q,T)=\alpha\,Q+\beta\,T$ \cite{Y.Xu et al}) and non-linear ($f(Q,T)=Q+\lambda\,Q^2+\eta\,T$ \cite{Dixit}) form and tried to find the exact WH solutions. After calculations, we found that it is not possible to find exact WH solutions analytically for the non-linear case. But for the linear case, it is possible. Hence, we consider
\begin{equation}
\label{2c}
f(Q,T)=\alpha\,Q+\beta\,T.
\end{equation}
 to study the wormhole solutions with linear EoS, where $\alpha$ and $\beta$ are free parameters.\\
Using the above linear functional form of $f(Q,T)$ in the field equations \eqref{17}-\eqref{19},  we obtain
 \begin{equation}\label{4a}
 \rho =\frac{\alpha  (12 \pi -\beta ) b'}{3 (4 \pi -\beta ) (\beta +8 \pi ) r^2},
 \end{equation}
 \begin{equation}\label{4b}
 p_r=-\frac{\alpha  \left(2 \beta  r b'-3 \beta  b+12 \pi  b\right)}{3 (4 \pi -\beta ) (\beta +8 \pi ) r^3},
 \end{equation}
 \begin{equation}\label{4c}
 p_t=-\frac{\alpha  \left((\beta +12 \pi ) r b'+3 b (\beta -4 \pi )\right)}{6 (4 \pi -\beta ) (\beta +8 \pi ) r^3}.
 \end{equation}
From Eqs. \eqref{4a} and \eqref{4b} by using the EoS given in Eq. \eqref{2a}  we obtain the shape function as
\begin{equation}\label{31}
b\left(r\right)= c r^\eta,
\end{equation}
where
\begin{equation}
\label{2d}
\eta={-\frac{3 (\beta -4 \pi )}{\beta  (\omega -2)-12 \pi  \omega }}\,\,,
\end{equation}
and $c$ is the integrating constant, and without loss of generality, one can consider $c=1$.\\
To satisfy the asymptotically flatness conditions, $\eta$ should be less than $1$, i.e., $\eta < 1$, and hence the possible ranges of $\omega$ and $\beta$ are listed below.\\
\begin{table}[H]
\begin{center}
\begin{tabular}{ |c|c|c| }
 \hline
 \multicolumn{2}{|c|}{Model Parameters} \\
 \hline
 $\omega$ & $\beta$ \\
 \hline
 $(-\infty, -1)$   & $(-\infty, \frac{12 \pi  \omega }{\omega -2})\cup (12 \pi, \infty)$\\
 \hline
$(-1, 2)$ &  $(\frac{12 \pi  \omega }{\omega -2}, 12 \pi)$ \\ 
\hline
 $2$ & $(-\infty, 12 \pi)$\\
 \hline
$(2, \infty)$ & $(-\infty, 12 \pi)\cup (\frac{12 \pi  \omega }{\omega -2}, \infty)$ \\
\hline
\end{tabular}
\caption{Possible domains of $\omega$ and $\beta$.}
\label{table:1}
\end{center}
\end{table}
From the above table, we can observe that $\omega<-1$ shows under the phantom region, and $-1<\omega<0$ shows under the quintessence region. As the universe is accelerating, we are neglecting the region $\omega\geq0$. By taking the range of $\omega$ and $\beta$  from the table into account, we fix some values and show the behavior of shape functions. Fig. \ref{Fig1} depicted the behavior of shape functions. One could observe that shape functions show positively increasing behavior under asymptotic background. Flaring out condition is also satisfied at the throat of the wormhole. In this case the throat radius $r_0=1$. Thus we could conclude that the shape function satisfies all the necessary conditions for its traversability.
Now, by taking the Eqs. \eqref{2c} and \eqref{31} into account, we are able to write the energy conditions below\\
$\bullet$ \textbf{NEC :} $\rho+p_r=-\frac{\alpha  (12 \pi -\beta ) (\omega +1) r^{\eta-3}}{\Lambda}$ and $\rho+p_t=\frac{\alpha \left( 12 \pi  (\omega -1)-\beta  (\omega -5)\right) r^{\eta-3}}{2\,\Lambda}$,\\
$\bullet$ \textbf{DEC :} $\rho-p_r=\frac{\alpha  (12 \pi -\beta )(\omega -1) r^{\eta-3}}{\Lambda}$ and $\rho-p_t=-\frac{\alpha\,\left( \beta\,(1-\omega ) +12 \pi  (\omega +3)\right) r^{\eta-3}}{2\,\Lambda}$,\\
$\bullet$ \textbf{SEC :} $\rho+p_r+2p_t=\frac{4\,\alpha\,\beta\,r^{\eta-3}}{\Lambda}$,\\
where, $\Lambda=(\beta +8 \pi ) \left(12 \pi  \omega -\beta  (\omega -2)\right)$.\\
Again, for this linear EoS case, one can find the the energy density from the field equation \eqref{20}
\begin{equation}
\label{2i}
\rho=\frac{\alpha  (\beta -12 \pi ) r^{\eta-3}}{\Lambda}.
\end{equation}
\begin{center}
\begin{figure}[h]
\includegraphics[width=7cm,height=4.5cm]{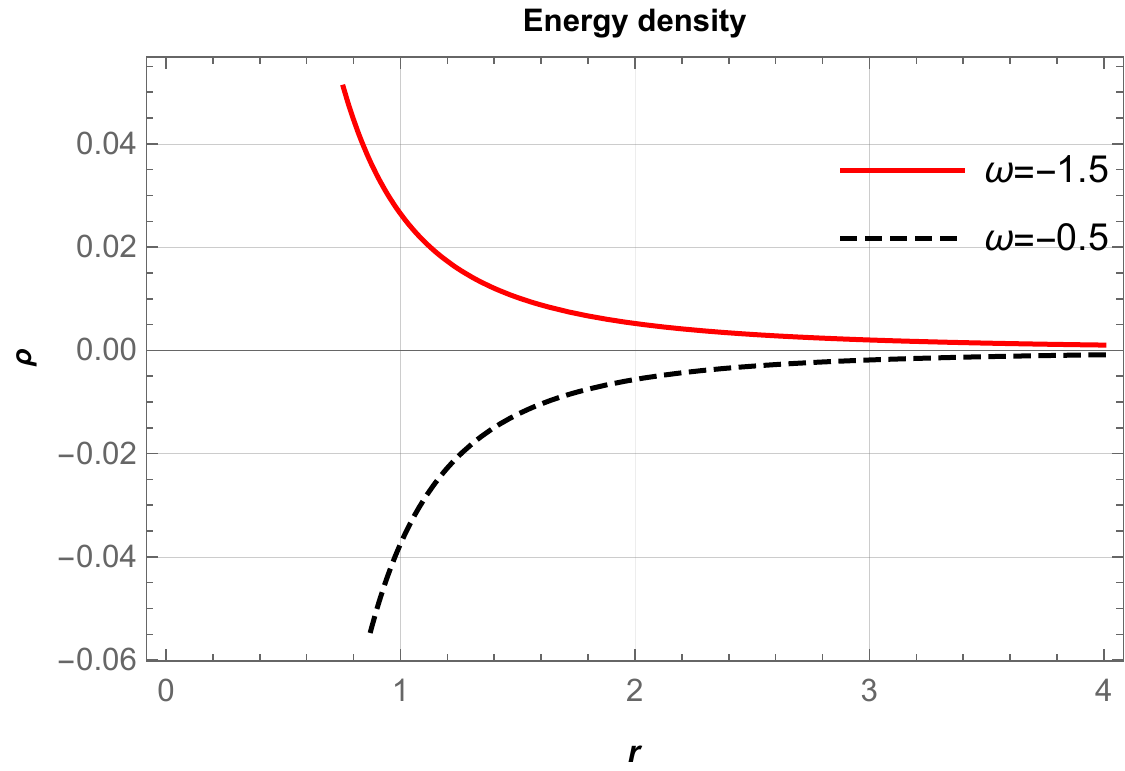}
\caption{The figure shows the behavior of energy density $\rho$ with respect to $r$ for particular values of $\omega=-1.5,-0.5$ corresponding to $\beta=1,14$, respectively. Here, we consider $\alpha=1$ and used the unit of radius in kilometers (km).}
\label{fig2}
\end{figure}
\end{center}
Considering some particular values of $\omega$ and $\beta$ based on the obtained range and plotted the graph for energy density in Fig. \ref{fig2}. Fig. \ref{fig2} depicted the graph for energy density $\rho$ for $\omega<-1$, which indicates that the energy density is showing positively decreasing behavior in the entire spacetime. But for $-1<\omega<0$, it is violating. Therefore, we have considered some particular value of $\omega\in(-\infty, -1)$ and plotted the graph for NEC, DEC, and SEC in Fig. \ref{fig3}. It can be observed from Fig. \ref{fig3} that NEC is violated for radial pressure and satisfied for tangential pressure in the entire spacetime. Also, DEC is satisfied with both pressures, whereas SEC is violated. Violation of NEC may confirm the presence of exotic matter at the throat of the wormhole.
\begin{table}[H]
\begin{center}
\begin{tabular}{ |c|c|c|c|c| }
 \hline
  \multicolumn{1}{|c|}{Terms} &
  \multicolumn{2}{|c|}{Interpretations}\\
 \hline
 $\omega$ & $(-\infty, -1)$ & $(-1, 2)$ \\
 \hline
 $\beta$   & $(-\infty, \frac{12 \pi  \omega }{\omega -2})\cup (12 \pi, \infty)$ & $(\frac{12 \pi  \omega }{\omega -2}, 12 \pi)$\\
 \hline
$\rho$ &  $satisfied$ & $violated$\\ 
 \hline
$\rho + p_r$ & $violated$ & $violated$\\
 \hline
$\rho + p_t$ & $satisfied$ & $satisfied$\\
 \hline
$\rho - p_r$ & $satisfied$ & $violated$\\
 \hline
$\rho - p_t$ & $satisfied$ & $violated$\\
 \hline
$\rho + p_r + 2p_t$ & $violated$ & $satisfied$\\
 \hline
\end{tabular}
\caption{Summary of the energy conditions for $p_r=\omega\,\rho$.}
\label{table:2}
\end{center}
\end{table}
\begin{figure}[H]
\centering
\includegraphics[width=15cm,height=13cm]{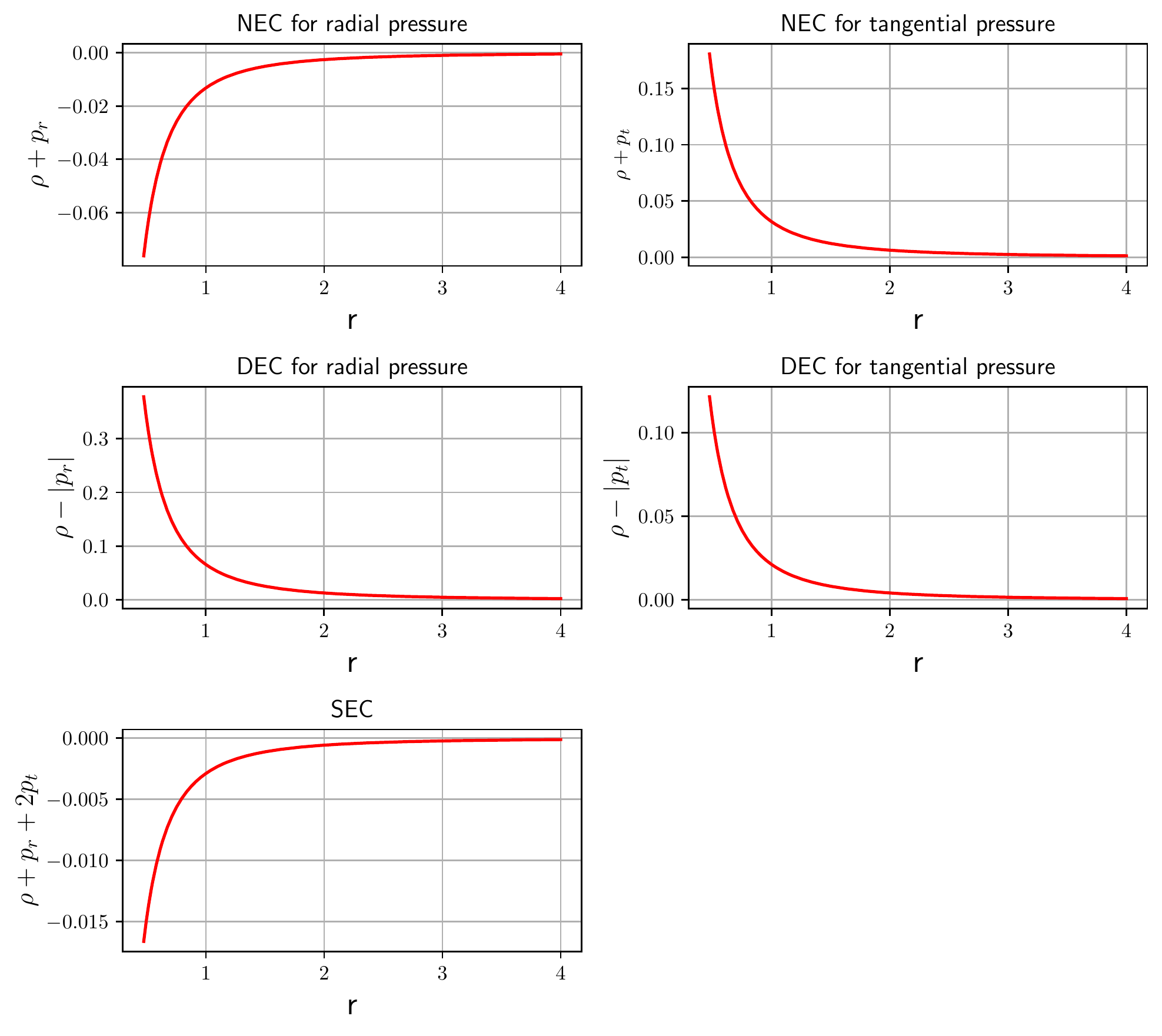}
\caption{Profile shows the behavior of NEC, DEC and SEC for $\omega=-1.5$ and $\beta=1$. Here we consider $\alpha=1$ and used the unit of radius in kilometers (km).}
\label{fig3}
\end{figure}
\subsubsection{\textbf{Anisotropic wormhole solutions}}
\begin{figure}[H]
\centering
\subfigure[]{\includegraphics[width=6.5cm,height=4cm]{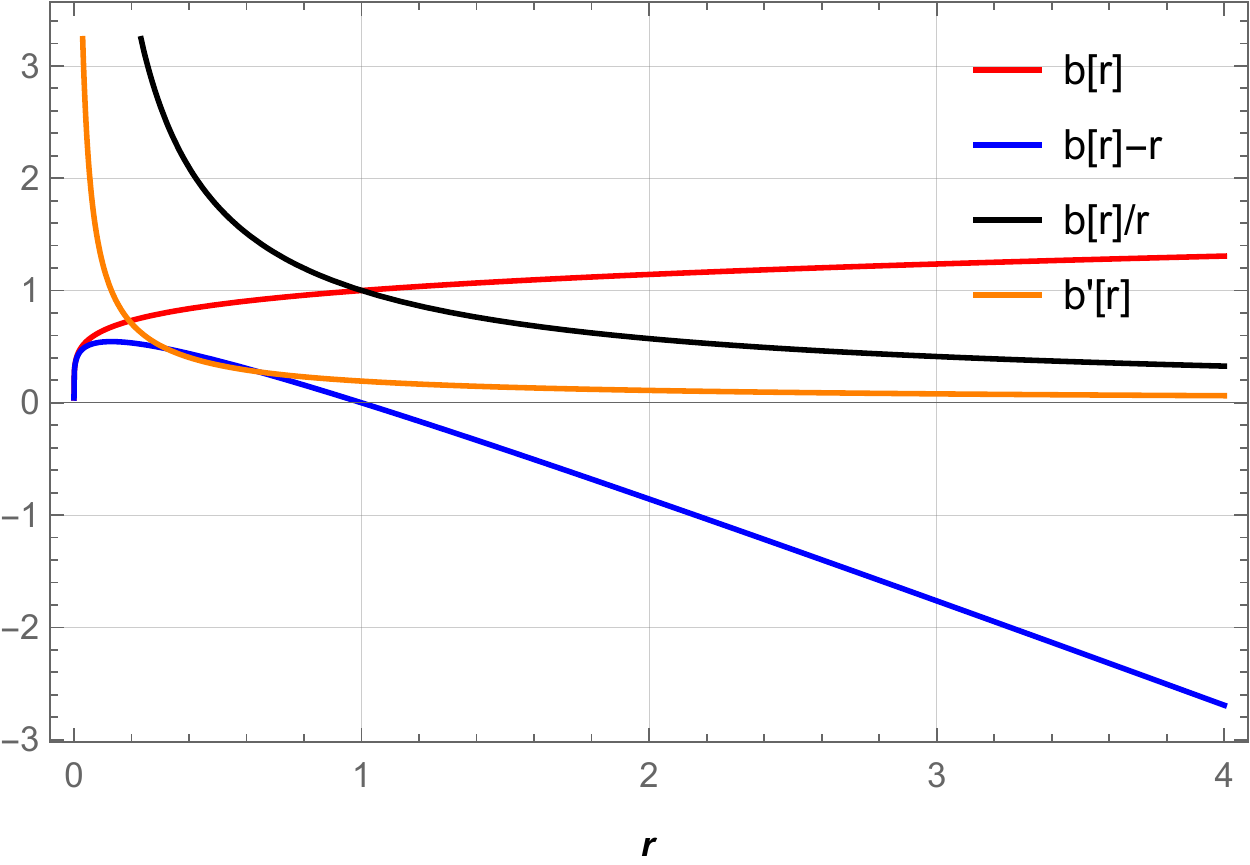}}\,\,\,\,\,\,\,\,\,\,\,\,\,\,\,\,\,\,\,\,\,\,\,\,\,
\subfigure[]{\includegraphics[width=6.5cm,height=4cm]{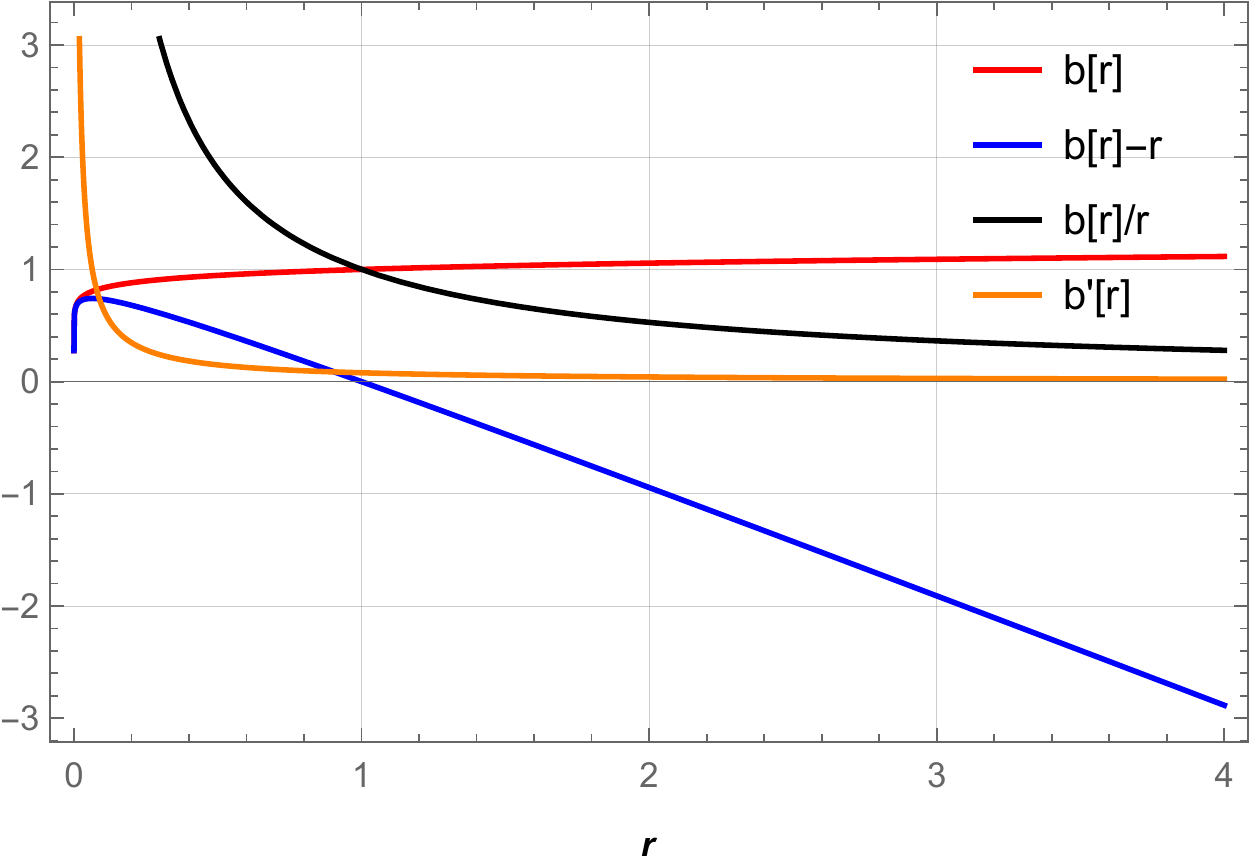}}
\caption{The behavior of shape function $b(r)$, flaring out condition $b'(r)<1$, throat condition $b(r)-r<0$ and assymptotically flatness condition $\frac{b(r)}{r}\rightarrow 0$ as $r\rightarrow \infty$ for (a) $n=-2.5$ with $\beta=16$ and (b) $n= -2$ with $\beta=14$. Also, we consider $\alpha=1$ and used the unit of radius in kilometers (km).}
\label{Fig6}
\end{figure}

\begin{figure}[H]
\centering
\subfigure[]{\includegraphics[width=6.5cm,height=4cm]{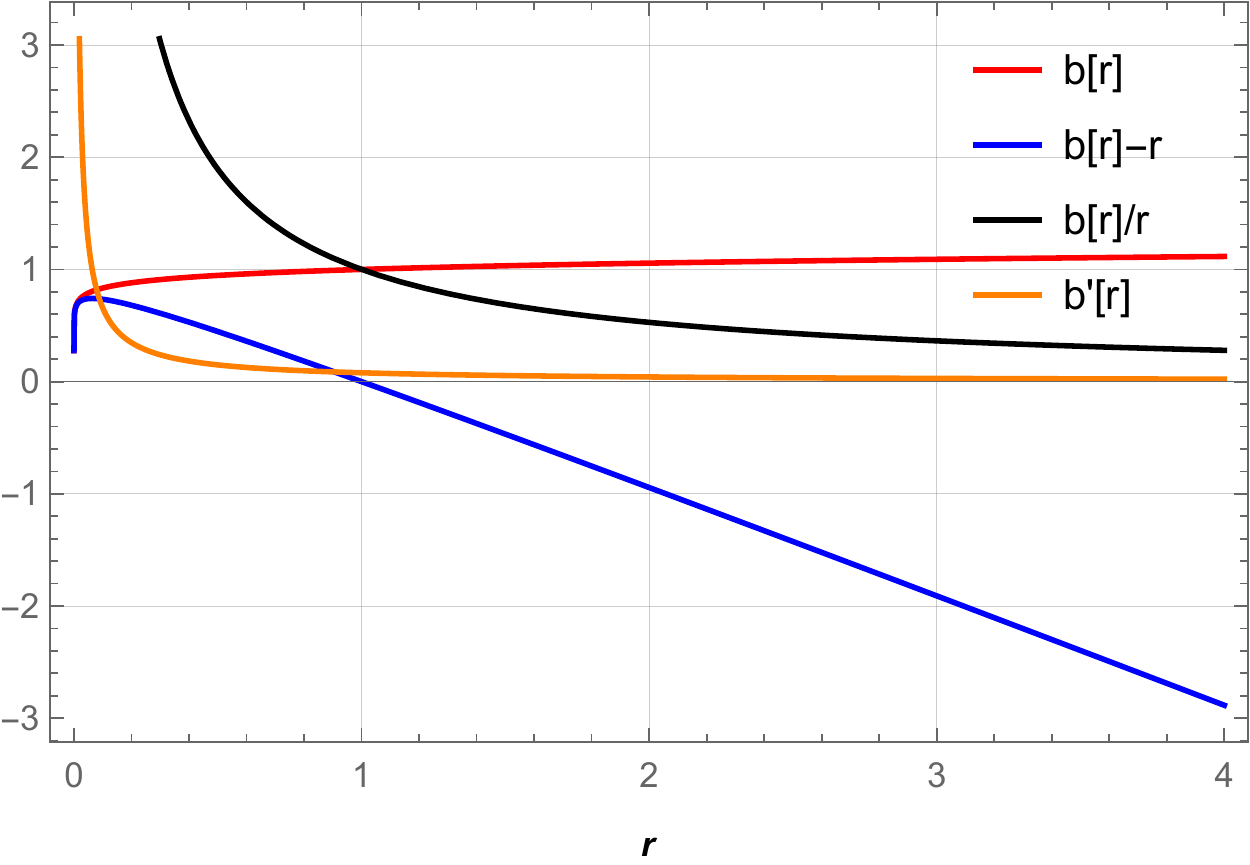}}\,\,\,\,\,\,\,\,\,\,\,\,\,\,\,\,\,\,\,\,\,\,\,\,\,
\subfigure[]{\includegraphics[width=6.5cm,height=4cm]{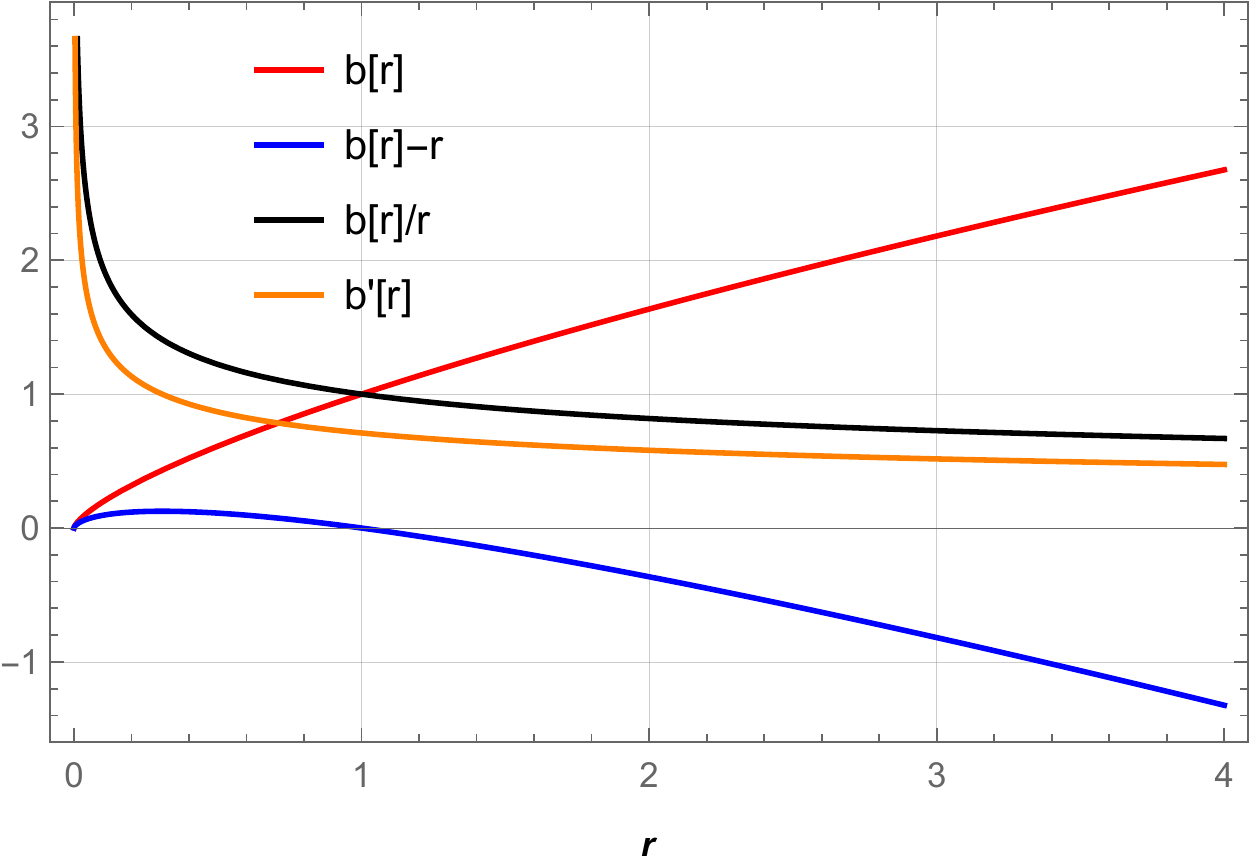}}
\caption{The behavior of shape function $b(r)$, flaring out condition $b'(r)<1$, throat condition $b(r)-r<0$ and assymptotically flatness condition $\frac{b(r)}{r}\rightarrow 0$ as $r\rightarrow \infty$ for (a) $n= -1.5$ with $\beta=14$ and (b) $n= -0.1$ with $\beta=1$. Also, we consider $\alpha=1$ and used the unit of radius in kilometers (km).}
\label{Fig7}
\end{figure}

\begin{figure}[H]
\centering
\subfigure[]{\includegraphics[width=6.5cm,height=4cm]{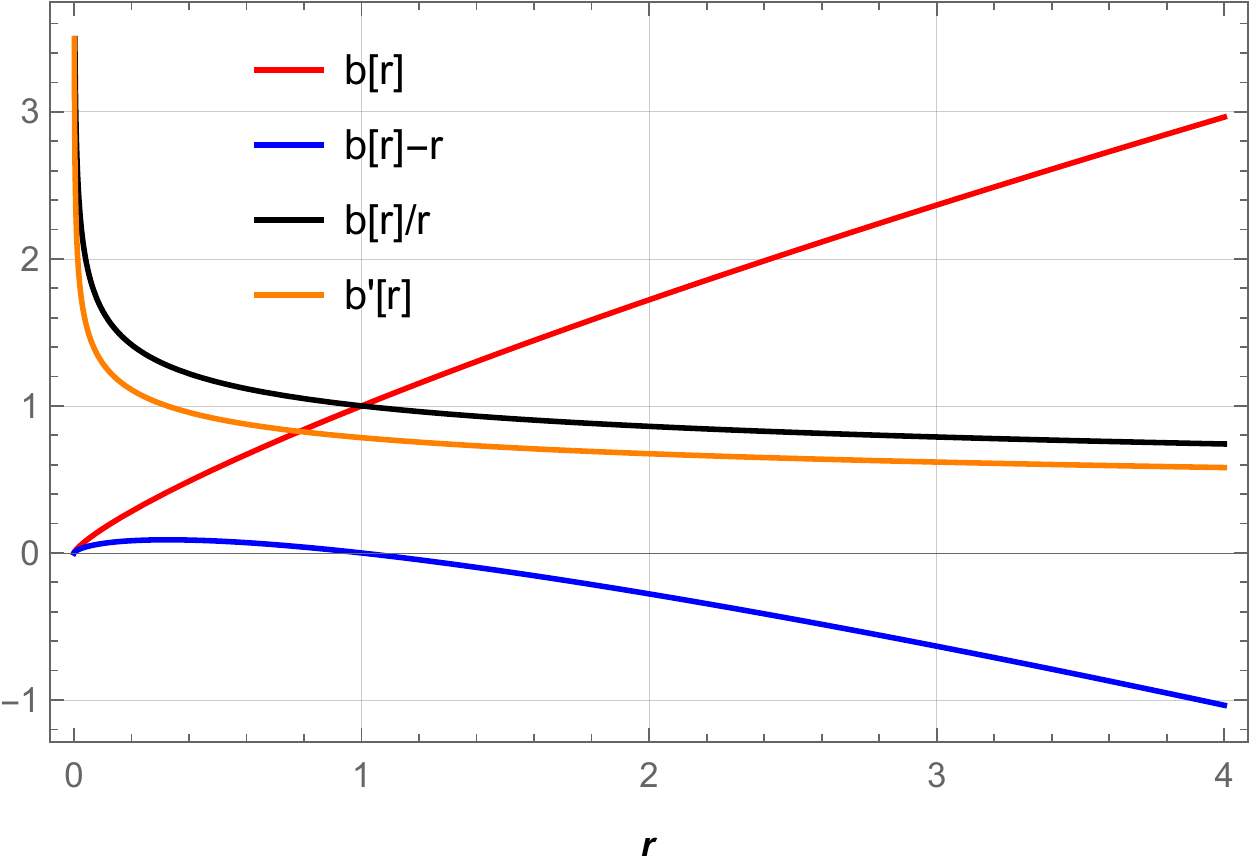}}\,\,\,\,\,\,\,\,\,\,\,\,\,\,\,\,\,\,\,\,\,\,\,\,\,
\subfigure[]{\includegraphics[width=6.5cm,height=4cm]{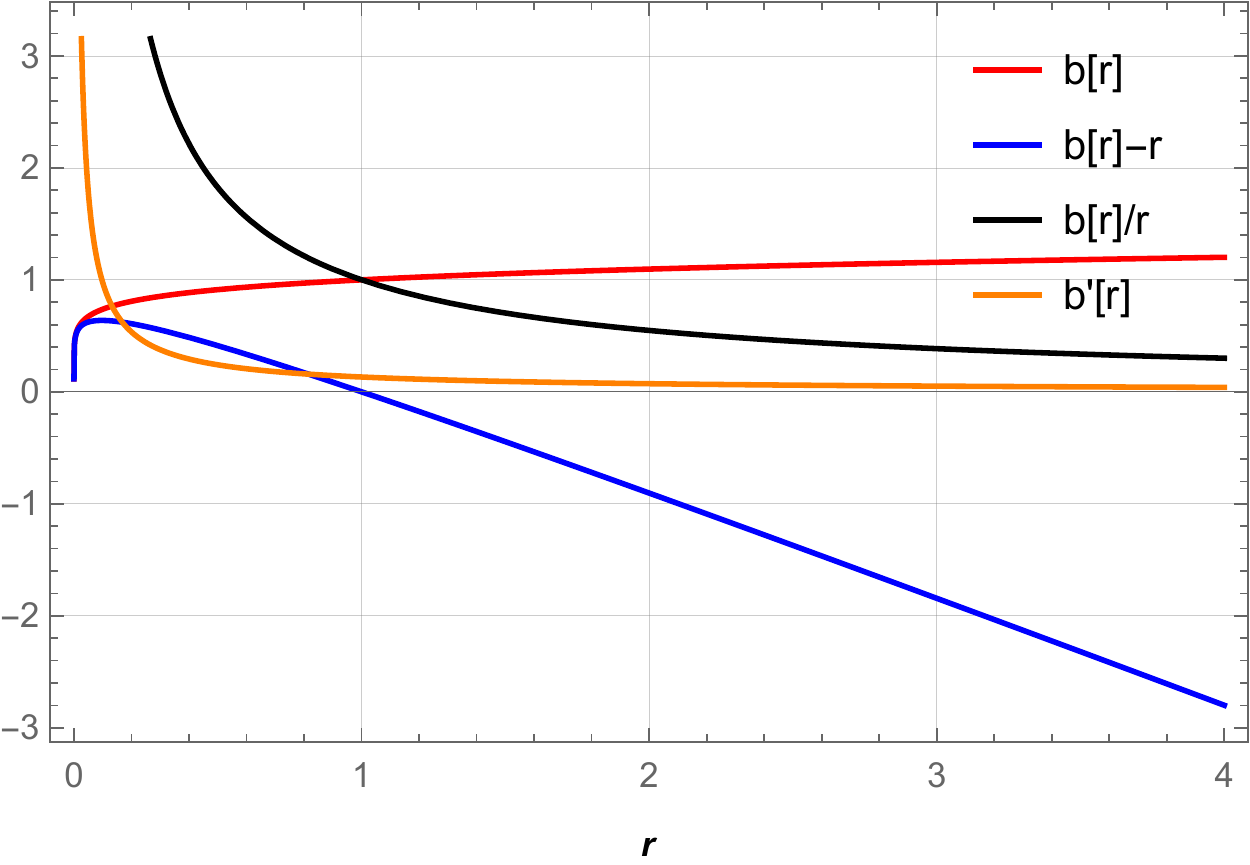}}
\caption{The behavior of shape function $b(r)$, flaring out condition $b'(r)<1$, throat condition $b(r)-r<0$ and asymptotically flatness condition $\frac{b(r)}{r}\rightarrow 0$ as $r\rightarrow \infty$ for (a) $n= 0.25$ with $\beta=6$ and (b) $n= 0.5$ with $\beta=12$. Also, we consider $\alpha=1$ and used the unit of radius in kilometers (km).}
\label{Fig8}
\end{figure}

\begin{figure}[H]
\centering
\includegraphics[width=6.5cm,height=4cm]{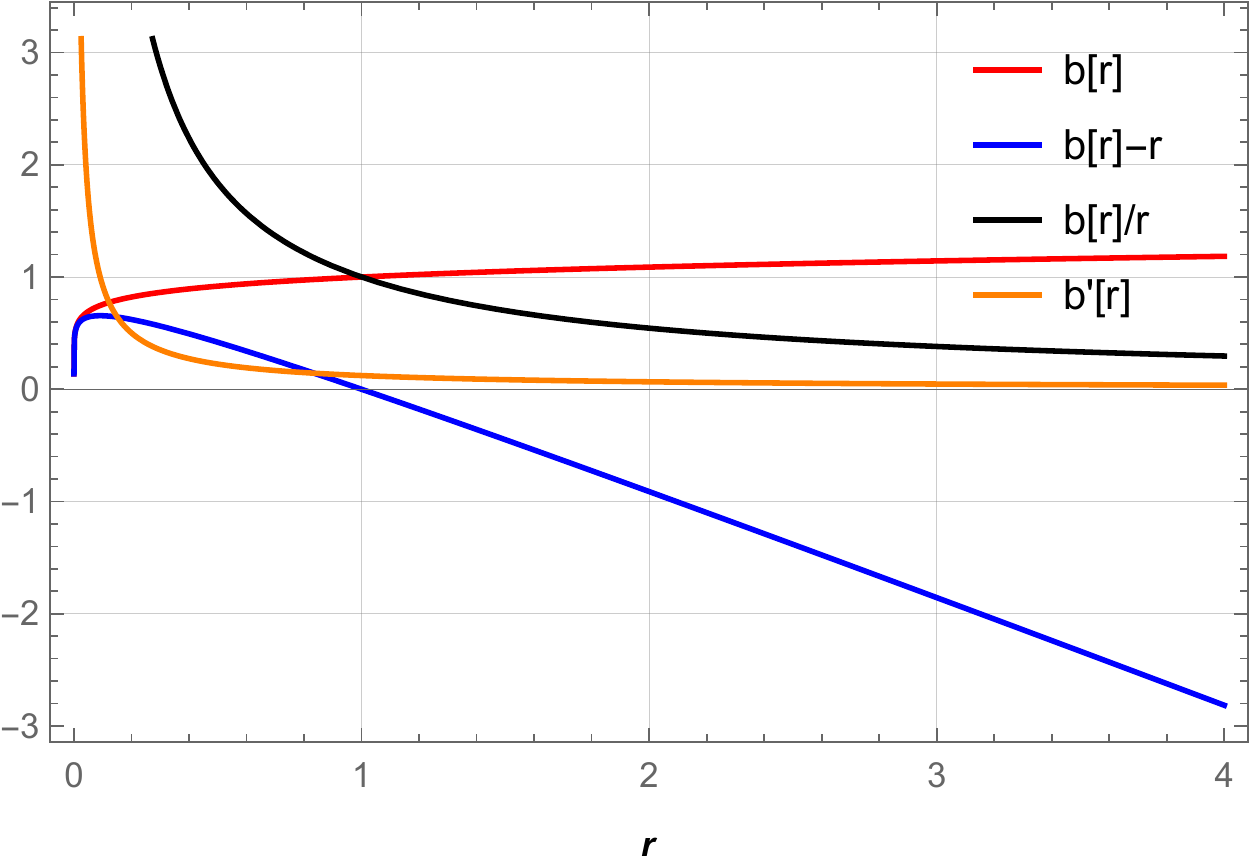}
\caption{The behavior of shape function $b(r)$, flaring out condition $b'(r)<1$, throat condition $b(r)-r<0$ and asymptotically flatness condition $\frac{b(r)}{r}\rightarrow 0$ as $r\rightarrow \infty$ for $n= 2$ with $\beta=13$. Also, we consider $\alpha=1$ and used the unit of radius in kilometers (km).}
\label{fig9}
\end{figure}

In this particular section, we have considered anisotropic energy momentum tensor fluid relation i.e. $p_r\neq p_t$ of the asymptotically flat wormhole solutions. For this case we consider the relation between $p_r$ and $p_t$ as follows\cite{Sarker,Moraes1}
\begin{equation}
\label{3a}
p_t = n\,p_r,
\end{equation}
where, $n$ is any parameter. Since we are studying with the anisotropic fluid, therefore $n$ can not be equal to one i.e., $n\neq1$, otherwise it will reduce to perfect fluid.\\
By using Eqs. \eqref{21}, \eqref{22} in Eq. \eqref{3a}, we will get
\begin{multline}
\label{3b}
f_Q \left[(n-1) f_T \left(3 b^2-r (b+2 r) b'\right)+24 \pi  \left(r b' (r-b n)+b (3 b n-2 n r-r)\right)\right]+\\
r (b-r) \left[2 b f_{\text{QQ}} Q' \left((n-1) f_T+24 \pi  n\right)+3 f (n-1) r^2 \left(f_T+8 \pi \right)\right]=0.
\end{multline}
For this case also we use the previous functional forms of $f(Q,T)$ and tried to find the shape function $b(r)$. We found that for non-linear $f(Q,T)=Q+\lambda\,Q^2+\eta\,T$, the exact wormhole solutions is formidable due to the complexity of field equations. But for linear $f(Q,T)=\alpha\,Q+\beta\,T$, we are able to get the exact solutions.\\
For the linear case, considering Eqs. \eqref{4b} and \eqref{4c} in Eq. \eqref{3a}, we get the shape function $b(r)$ as follows
\begin{equation}
\label{3c}
b(r)=c_1 r^{\gamma},
\end{equation}
where
\begin{equation}
\label{3d}
\gamma=\frac{3 (4 \pi -\beta ) (2 n+1)}{\beta -4 \beta  n+12 \pi },
\end{equation}
and $c_1$ is the integrating constant and without loss of generality one  can consider $c_1=1$.\\
To satisfy the asymptotically flatness conditions, $\gamma$ should be less than $1$ i.e., $\gamma < 1$ and hence the possible ranges of $n$ and $\beta$ are listed below
\begin{table}[H]
\begin{center}
\begin{tabular}{ |c|c|c| }
 \hline
 \multicolumn{2}{|c|}{Model Parameters} \\
 \hline 
 $n$ & $\beta$ \\
 \hline
 $(-\infty, -2)$   & $(\frac{12 \pi}{4n-1}, \frac{12 \pi n}{n+2})$\\
 \hline
 $-2$ & $(\frac{-4 \pi}{3}, \infty)$ \\
\hline
$(-2, \frac{-1}{2}]$ & $(-\infty, \frac{12 \pi n}{n+2}))\cup (\frac{12 \pi}{4n-1}, \infty)$ \\
\hline
$(\frac{-1}{2}, \frac{1}{4})$ & $(-\infty, \frac{12 \pi}{4n-1}) \cup (\frac{12 \pi n}{n+2}, \infty)$  \\
\hline
$\frac{1}{4}$ & $(\frac{4 \pi}{3}, \infty)$ \\
\hline
$(\frac{1}{4}, 1)$ & $(\frac{12 \pi n}{n+2}, \frac{12 \pi}{4n-1})$ \\
\hline
$(1, \infty)$ & $(\frac{12 \pi}{4n-1}, \frac{12 \pi n}{n+2})$  \\
\hline
\end{tabular}
\caption{Possible domains of $\beta$ and $n$.}
\label{table:3}
\end{center}
\end{table}
From the above table, we picked up some particular values of $\beta$ and $n$ from each domain and plotted the graph for shape functions in Figs. \ref{Fig6}-\ref{fig9}. It is clear from the figures that the shape functions satisfy all the necessary conditions for a traversable wormhole. In this case, we get the wormhole's throat radius $r_0=1$.\\
Again, by considering the Eqs. \eqref{2c} and \eqref{3c}, we obtain the energy conditions which are defined below\\
$\bullet$ \textbf{NEC :} $\rho+p_r=\frac{r^{k} (24 \pi  \alpha   n-2 \alpha  \beta   (n+2))}{\Lambda_{1}}$ and $\rho+p_t=\frac{\alpha   (12 \pi  (n+1)-\beta  (5 n+1)) r^{k}}{\Lambda_{1}}$,\\
$\bullet$ \textbf{DEC :} $\rho-p_r=\frac{2 \alpha   (\beta -\beta  n+12 \pi  (n+1)) r^{k}}{\Lambda_{1}}$ and $\rho-p_t=\frac{\alpha   (\beta  (n-1)+12 \pi  (3 n+1)) r^{k}}{\Lambda_{1}}$,\\
$\bullet$ \textbf{SEC :} $\rho+p_r+2p_t=-\frac{4 \alpha  \beta   (2 n+1) r^{k}}{\Lambda_{1}}$,\\
where $\Lambda_{1}=(\beta +8 \pi ) (\beta -4 \beta  n+12 \pi )$ and $k=-\frac{6 (\beta +4 \pi ) (n-1)}{\beta  (4 n-1)-12 \pi }$.\\
Again, for this case, one can find the the energy density from the field equation \eqref{20}
\begin{equation}
\label{2j}
\rho=\frac{\alpha  (12 \pi -\beta )  (2 n+1) r^k}{\Lambda_1}.
\end{equation}
\begin{figure}[H]
\centering
\includegraphics[width=7.5cm,height=5cm]{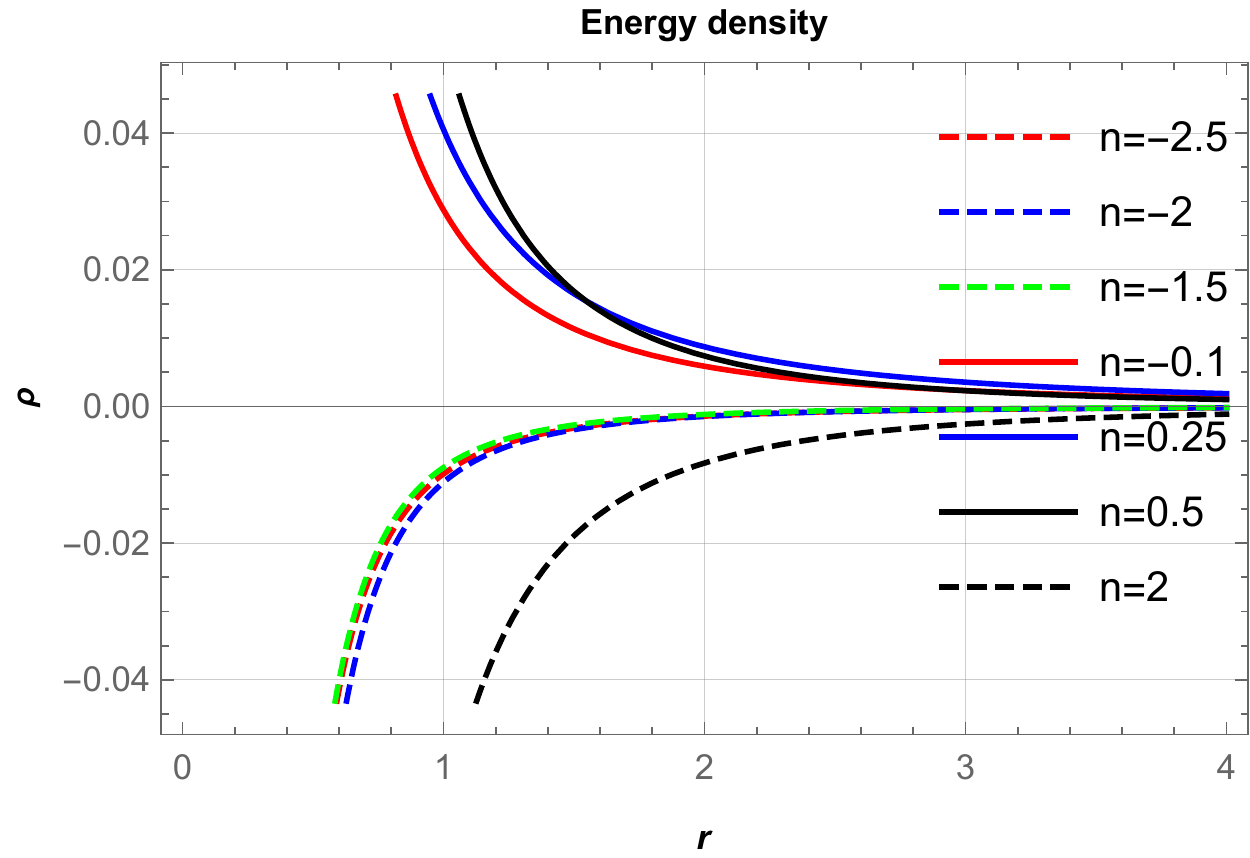}
\centering
\caption{The figure shows the behavior of $\rho$ with respect to $r$ for particular value of $n=-2.5,\,-2,\,-1.5,\,-0.1,\,0.25,\,0.5,$ and $2$ corresponding to $\beta=16,\,14,\,14,\,1,\,6,\,12,$ and $13$, respectively. We consider $\alpha=1$ and used the unit of radius in kilometers (km).}
\label{fig10}
\end{figure}
Now considering some particular values from each domain from the table-\ref{table:3} and plotted the graph for energy density which is depicted in  Fig. \ref{fig10}. One can check that energy density is positive for $\frac{-1}{2}<n<\frac{1}{4}$, $n=\frac{1}{4}$, $\frac{1}{4}<n<1$. But for other region of $n$, it is violating. Therefore, we have considered some particular values of $n$ from $\frac{-1}{2}<n<\frac{1}{4}$, $n=\frac{1}{4}$, $\frac{1}{4}<n<1$ and plotted the graph for NEC, DEC, and SEC in Figs. \ref{fig11}-\ref{fig13}. It can be observed from Fig. \ref{fig11} that NEC is violated for radial pressure and satisfied for tangential pressure in the entire spacetime. Moreover, from Figs. \ref{fig12} and \ref{fig13} we could conclude that DEC is satisfied for both pressures while SEC is violated.
\begin{figure}[H]
\centering
\includegraphics[width=6.5cm,height=4cm]{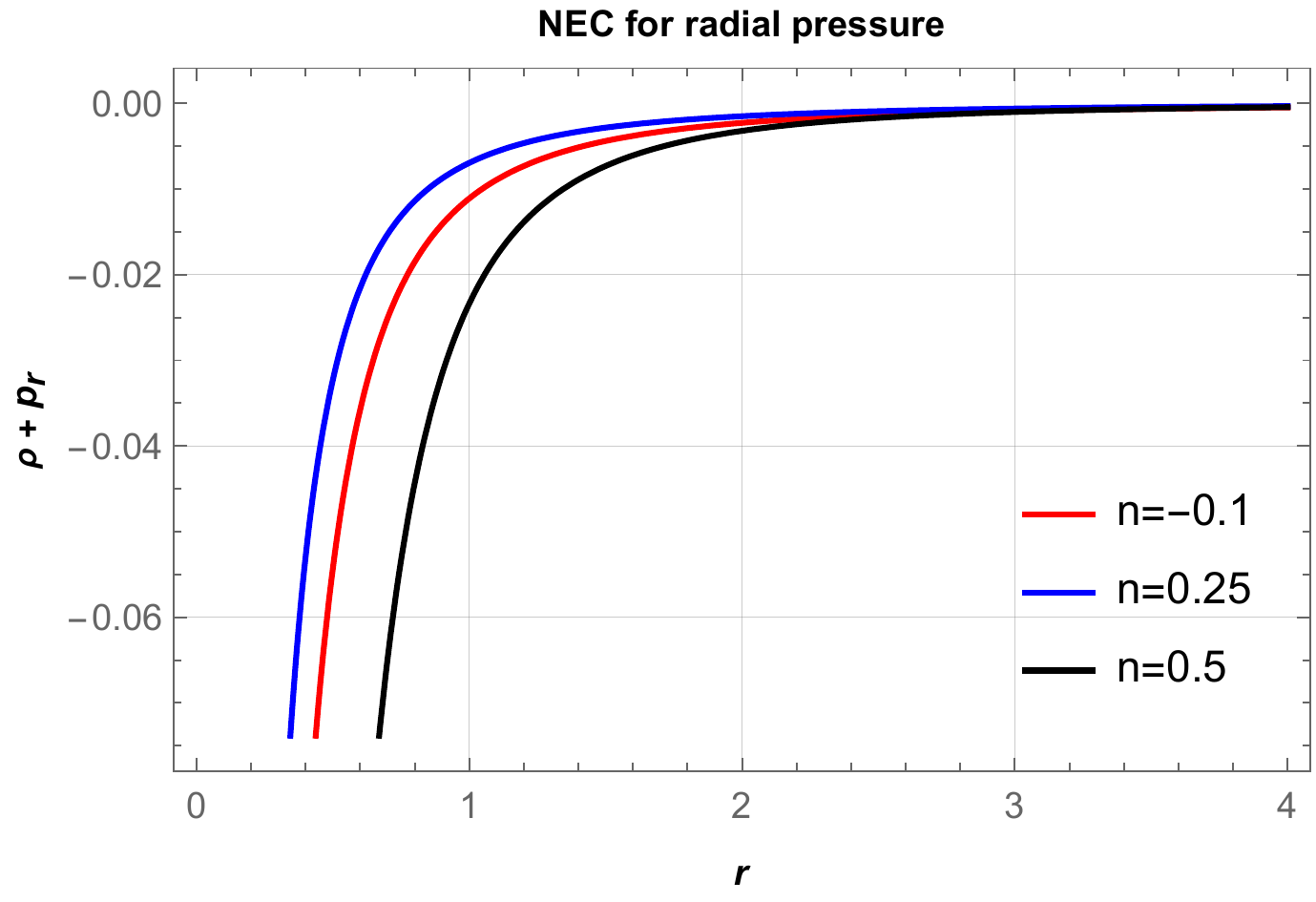}\,\,\,\,\,\,\,\,\,\,\,\,\,\,\,\,\,\,\,\,\,\,\,\,\,
\includegraphics[width=6.5cm,height=4cm]{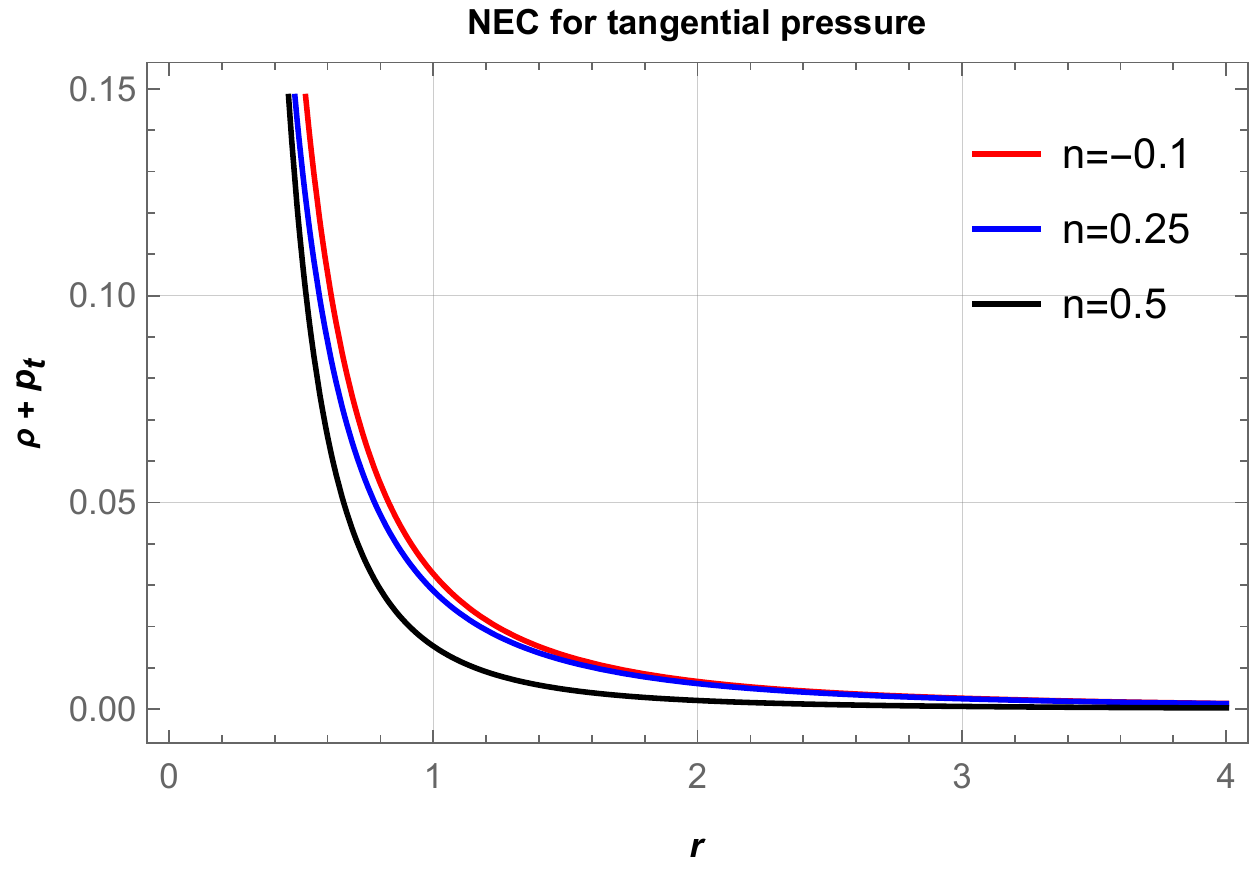}
\caption{Profile shows the behavior of NEC for $n=-0.1,\,0.25,$ and $0.5$ corresponding to $\beta=1,\,6,$ and $12$, respectively. We consider $\alpha=1$ and used the unit of radius in kilometers (km).}
\label{fig11}
\end{figure}
\begin{figure}[H]
\centering
\includegraphics[width=6.5cm,height=4cm]{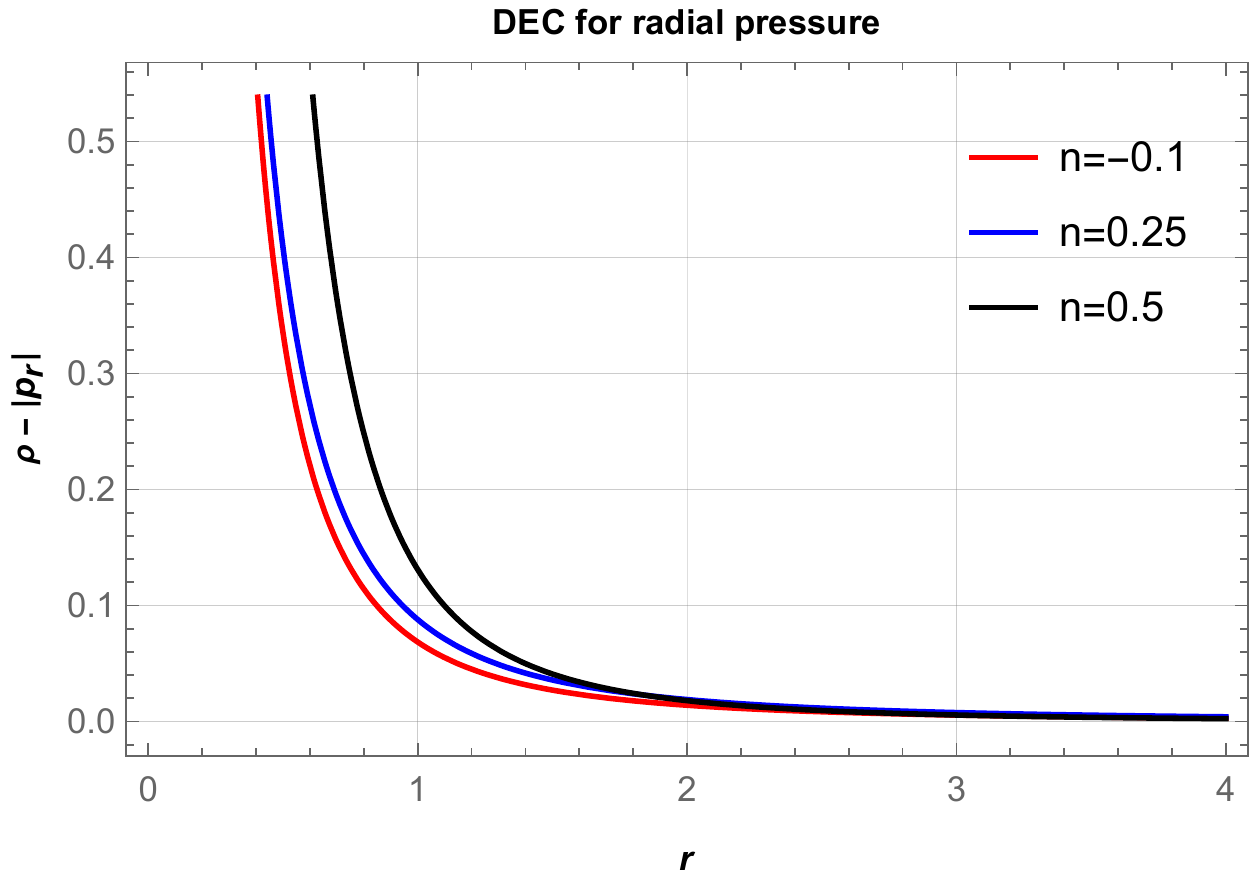}\,\,\,\,\,\,\,\,\,\,\,\,\,\,\,\,\,\,\,\,\,\,\,\,\,
\includegraphics[width=6.5cm,height=4cm]{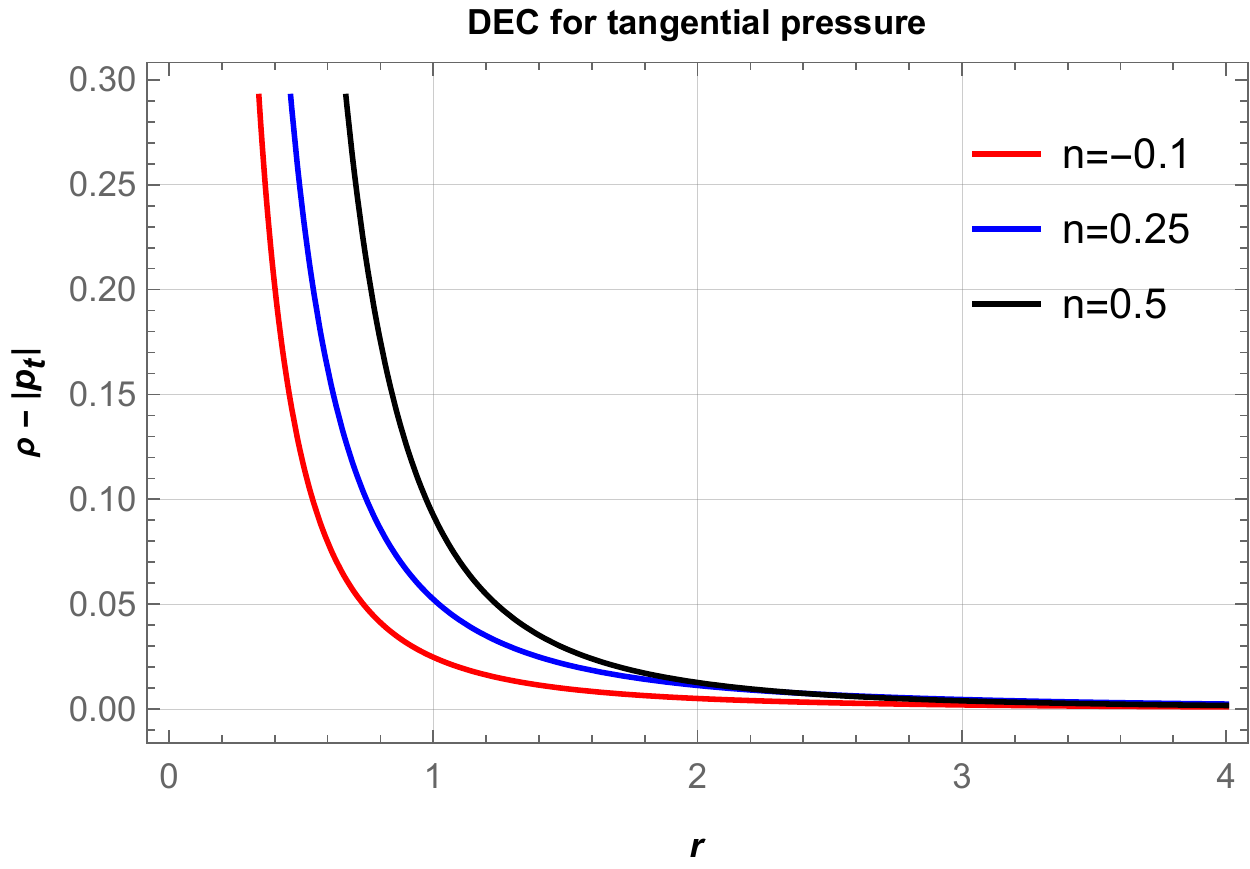}
\caption{Profile shows the behavior of DEC for $n=-0.1,\,0.25,$ and $0.5$ corresponding to $\beta=1,\,6,$ and $12$, respectively. We consider $\alpha=1$ and used the unit of radius in kilometers (km).}
\label{fig12}
\end{figure}
\begin{figure}[H]
\centering
\includegraphics[width=6.5cm,height=4cm]{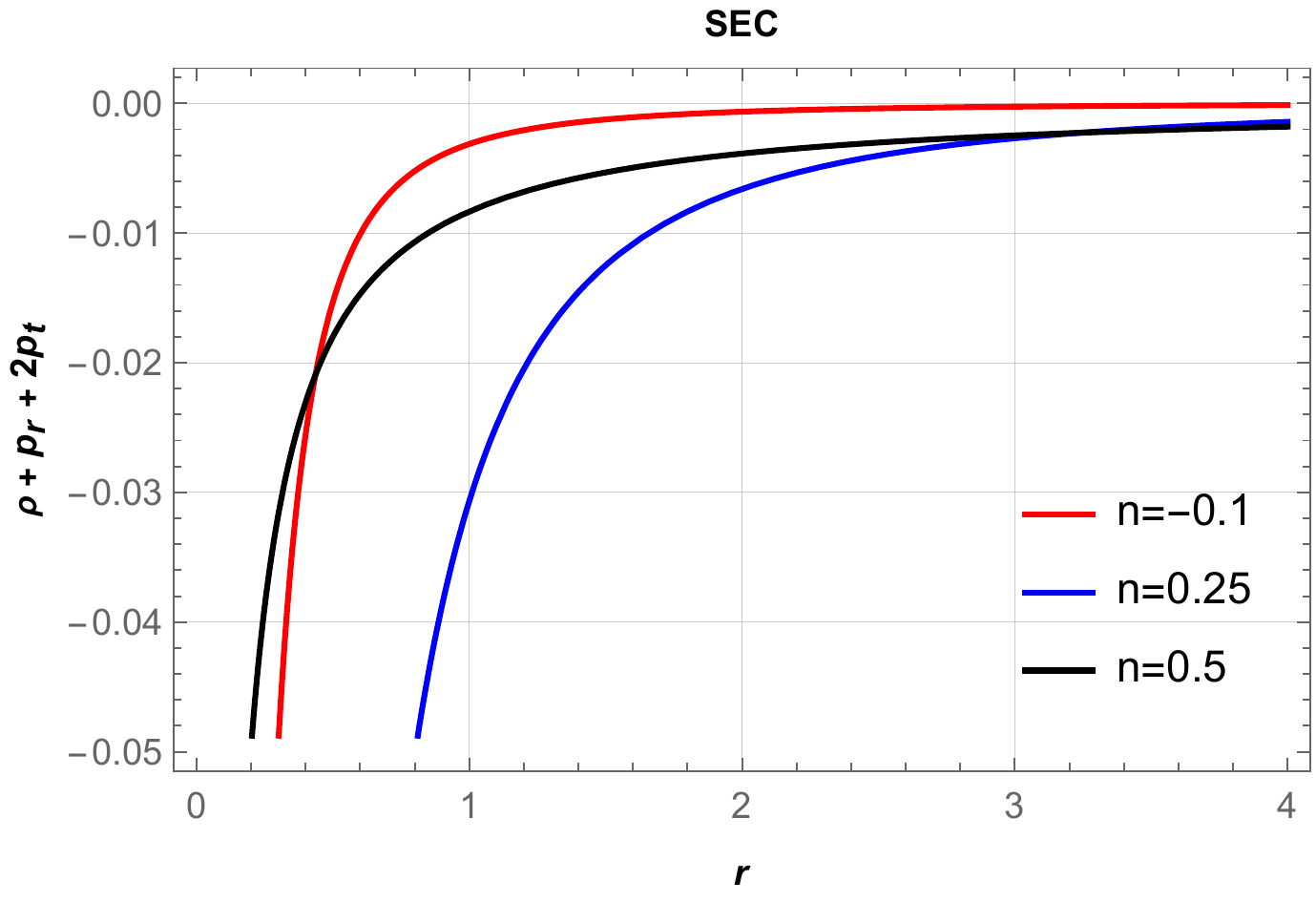}
\centering
\caption{Profile shows the behavior of SEC for $n=-0.1,\,0.25,$ and $0.5$ corresponding to $\beta=1,\,6,$ and $12$, respectively. We consider $\alpha=1$ and used the unit of radius in kilometers (km).}
\label{fig13}
\end{figure}
\begin{table}[H]
\begin{center}
\begin{tabular}{ |c|c|c|p{2cm}|p{2cm}|c|c|c| }
 \hline
  \multicolumn{1}{|c|}{Terms} &
  \multicolumn{7}{|c|}{Interpretations}\\
 \hline
 $n$ & $(-\infty, -2)$ & $-2$ & $(-2, \frac{-1}{2}]$ & $(\frac{-1}{2}, \frac{1}{4})$ & $\frac{1}{4}$ & $(\frac{1}{4}, 1)$ & $(1, \infty)$ \\
 \hline
 $\beta$   & $(\frac{12 \pi}{4n-1}, \frac{12 \pi n}{n+2})$ & $(\frac{-4 \pi}{3}, \infty)$ & $(-\infty, \frac{12 \pi n}{n+2})\cup (\frac{12 \pi}{4n-1}, \infty)$ & $(-\infty, \frac{12 \pi}{4n-1}) \cup (\frac{12 \pi n}{n+2}, \infty)$ & $(\frac{4 \pi}{3}, \infty)$ & $(\frac{12 \pi n}{n+2}, \frac{12 \pi}{4n-1})$ & $(\frac{12 \pi}{4n-1}, \frac{12 \pi n}{n+2})$\\
 \hline
$\rho$ &  $violated$ & $violated$ & $violated$ & $satisfied$ & $satisfied$ & $satisfied$ & $violated$\\ 
 \hline
$\rho + p_r$ &  $violated$ & $violated$ & $violated$ & $violated$ & $violated$ & $violated$ & $violated$\\
 \hline
$\rho + p_t$ &  $satisfied$ & $satisfied$ & $satisfied$ & $satisfied$ & $satisfied$ & $satisfied$ & $satisfied$\\
 \hline
$\rho - p_r$ &  $violated$ & $satisfied$ & $satisfied$ & $satisfied$ & $satisfied$ & $satisfied$ & $violated$\\
 \hline
$\rho - p_t$ &  $violated$ & $violated$ & $violated$ & $satisfied$ & $satisfied$ & $satisfied$ & $violated$\\
 \hline
$\rho + p_r + 2p_t$ &  $satisfied$ & $satisfied$ & $satisfied$ & $violated$ & $violated$ & $violated$ & $satisfied$\\
 \hline
\end{tabular}
\caption{Summary of the energy conditions for $p_t=n\,p_r$.}
\label{table:4}
\end{center}
\end{table}
\section{Equilibrium Conditions}
\label{sec5}
In this section, we consider the generalized Tolman-Oppenheimer-Volkov (TOV) equation \cite{Oppenheimer,Gorini,Kuhfittig} to find the stability of our obtained wormhole solutions. The generalized TOV equation can be written as
\begin{eqnarray}\label{51}
\frac{\varpi^{'}}{2}(\rho+p_r)+\frac{dp_r}{dr}+\frac{2}{r}(p_r-p_t)=0,
\end{eqnarray}
where $\varpi=2\phi(r)$.\\
Due to anisotropic matter distribution, the hydrostatic, gravitational, and anisotropic forces are defined as follows
\begin{eqnarray}\label{52}
F_h=-\frac{dp_r}{dr}, ~~~~~F_g=-\frac{\varpi^{'}}{2}(\rho+p_r), ~~~~F_a=\frac{2}{r}(p_t-p_r).
\end{eqnarray}
To equilibrium the wormhole solutions, it is necessary that $F_h+F_g+F_a=0$ must hold. Since in this study, we have assumed the redshift function $\phi(r)=constant$, so it will vanish the gravitational contribution $F_g$ in the equilibrium equation. Hence, the equilibrium equation become
\begin{equation}
\label{52a}
F_h+F_a=0.
\end{equation}
Using Eqs. \eqref{21},\eqref{22},\eqref{2a} and \eqref{2c}, we get the following equations for the hydrostatic and anisotropic forces for the linear model using $p_r=\omega \rho$ as
\begin{equation}
F_h=-\frac{(\beta -12 \pi )  \omega  \left(\frac{3 (\beta -4 \pi )}{12 \pi  \omega -\beta  (\omega -2)}-3\right) r^{\frac{3 (\beta -4 \pi )}{12 \pi  \omega -\beta  (\omega -2)}-4}}{(\beta +8 \pi ) (12 \pi  \omega -\beta  (\omega -2))},
\end{equation}
\begin{equation}
F_a=\frac{3 \alpha   (\beta  (-\omega )+\beta +4 \pi  (3 \omega +1)) r^{\frac{3 (\beta -4 \pi )}{12 \pi  \omega -\beta  (\omega -2)}-4}}{(\beta +8 \pi ) (12 \pi  \omega -\beta  (\omega -2))}.
\end{equation}
Also for the relation $p_t=n p_r$, considering Eqs. \eqref{21},\eqref{22}, \eqref{3a} and \eqref{2c}, the equations for the hydrostatic and anisotropic forces reads as
\begin{equation}
F_h=-\frac{18 \alpha  (\beta +4 \pi )^2  (n-1) r^{-\frac{6 (\beta +4 \pi ) (n-1)}{\beta  (4 n-1)-12 \pi }-1}}{(\beta +8 \pi ) (\beta -4 \beta  n+12 \pi ) (\beta  (4 n-1)-12 \pi )},
\end{equation}
\begin{equation}
F_a=-\frac{6 \alpha  (\beta +4 \pi )  (n-1) r^{-\frac{6 (\beta +4 \pi ) (n-1)}{\beta  (4 n-1)-12 \pi }-1}}{(\beta +8 \pi ) (\beta -4 \beta  n+12 \pi )}.
\end{equation}
\begin{figure}[H]
\centering
\includegraphics[width=6.5cm,height=4cm]{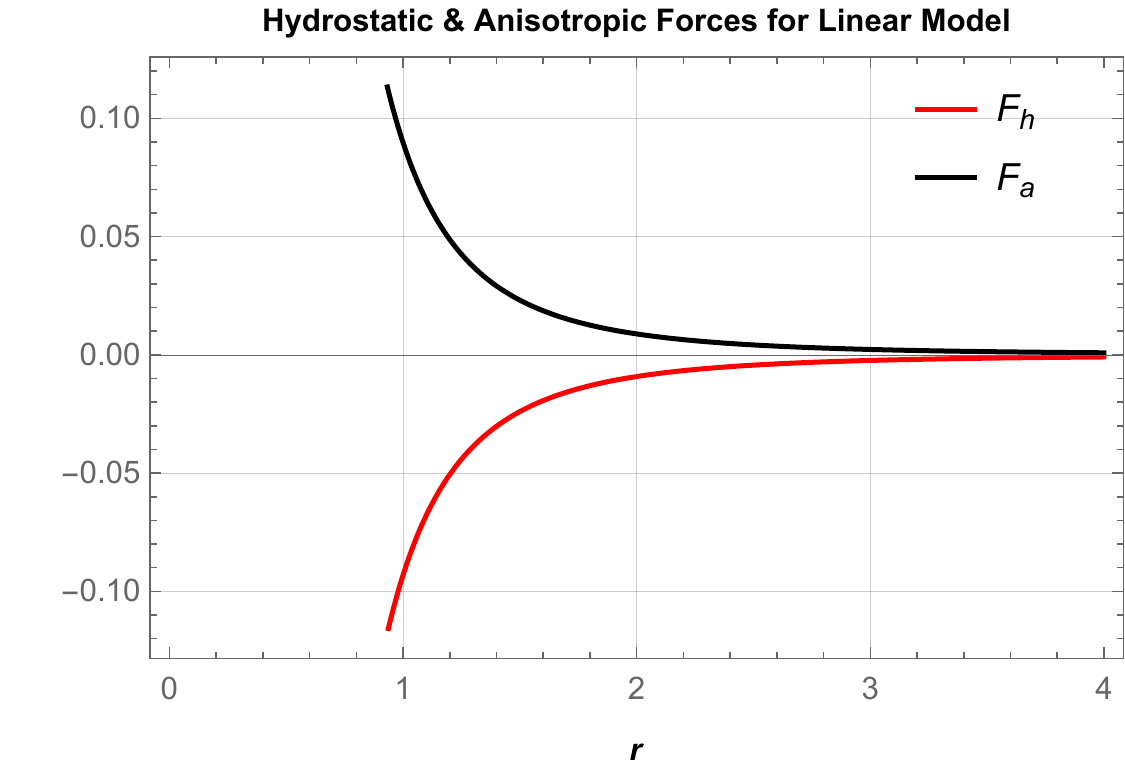}
\caption{This profile shows the behavior of both hydrostatic and anisotropic forces for $p_r=\omega \rho$ with $\omega=-1.5$. Also, we consider $\alpha=1$ and used the unit of radius in kilometers (km).}
\label{fig14}
\end{figure}
\begin{figure}[H]
\includegraphics[width=6.5cm,height=4cm]{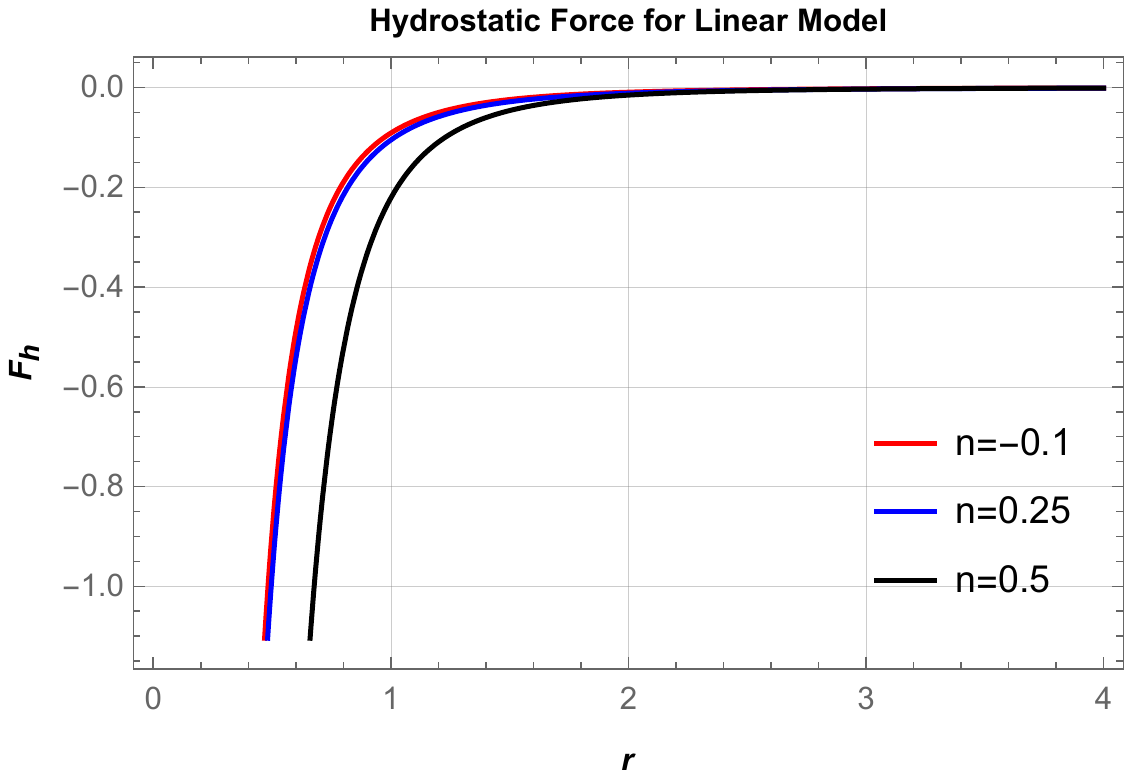}\,\,\,\,\,\,\,\,\,\,\,\,\,\,\,\,\,\,\,\,\,\,\,\,\,
\includegraphics[width=6.5cm,height=4cm]{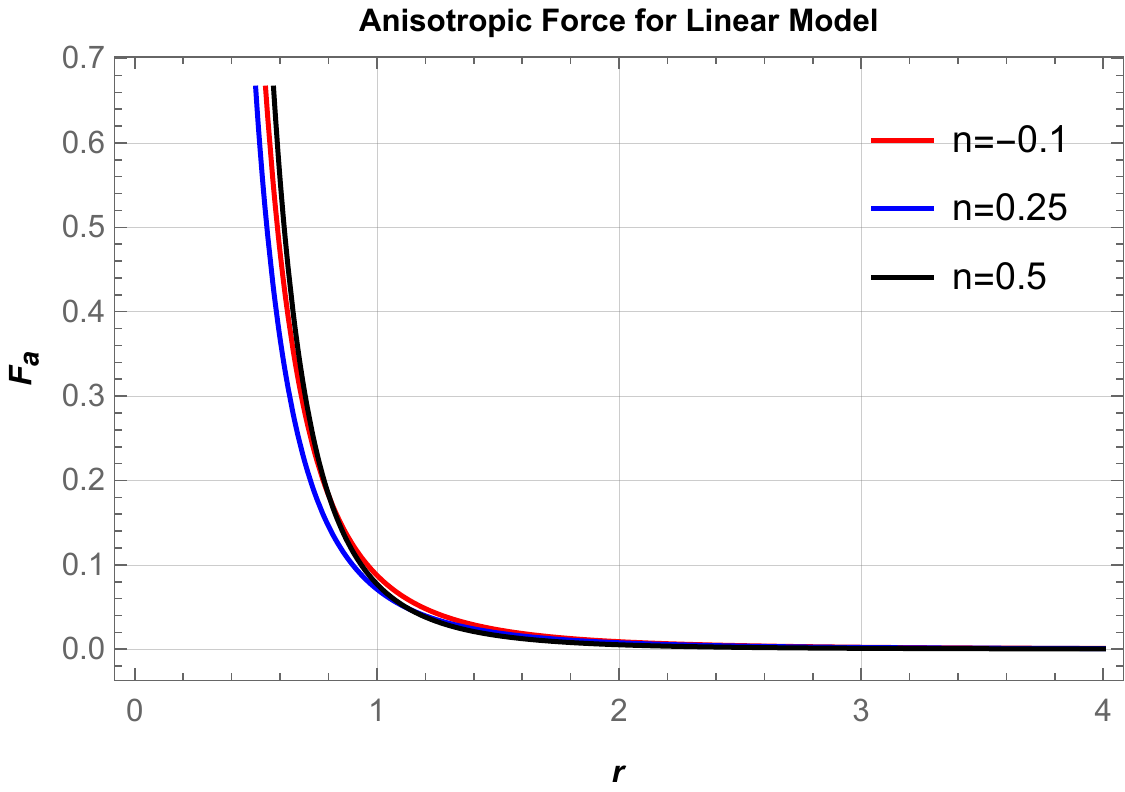}\\
\centering
\includegraphics[width=6.5cm,height=4cm]{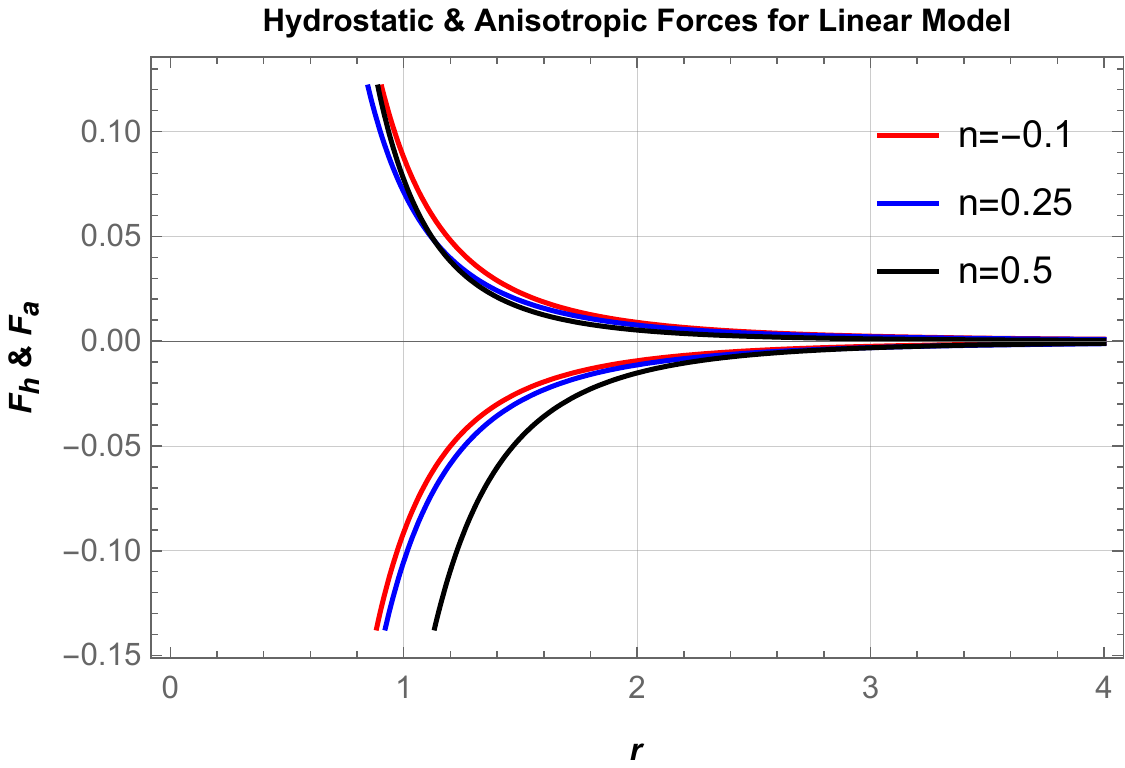}
\caption{This profile shows the behavior of both hydrostatic and anisotropic forces for $p_t=n p_r$. Also, we consider $\alpha=1$ and used the unit of radius in kilometers (km).}
\label{fig15}
\end{figure}
The graphs of hydrostatic and anisotropic forces for both cases are depicted in Figs. \ref{fig14} and \ref{fig15}. It can be observed that these forces show the same behavior but are opposite to each other. These balancing developments indicate that our obtained wormhole solutions are stable.
%

\section{Final Remarks}
\label{sec6}
Wormholes can act as tunnels that connects two different spatially  regions separated by a spacelike interval of the same spacetime manifold. It is as of now a theoretical possibility that has not been observed yet. Based on the advances in gravity wave astronomy, one may be able to conjecture a scenario where the signatures of astrophysical wormhole may be detected. In GR, these wormhole solutions that are of interest (stable, traversable) suffer from a known pathology where energy conditions are violated. The requirement for exotic matter where one explores standard model extensions in the context of classical general relativity is a big hurdle. One can now turn to modified versions of gravity to solve the issue regarding the wormholes. The study of wormholes in modified theory of gravity makes it possible to derive stable solutions without compromising the energy conditions.\\
The modified theory of gravity, $f(Q,T)$ has successfully explained the late-time acceleration \cite{Arora1} and the matter-antimatter asymmetry \cite{Bhattacharjee}. It is a worthwhile question to ask whether modified theories of gravity are useful in the context of astrophysical objects like a wormhole and if true, then which modified versions support stable, traversible wormholes. This is the motivation to study wormhole solutions in recently proposed $f(Q,T)$ gravity. Since $f(Q,T)$ gravity is a novel approach, it may provide new insights into astrophysical objects like blackholes and wormholes. In our work, we  set up the corresponding field equations in $f(Q,T)$ gravity. We then  conduct our analysis in two phases: (i) wormhole solutions with a linear EoS and (ii) anisotropic wormhole solutions. Further, we study these solutions under two functional forms of the $f(Q,T)$ models. We have considered linear $f(Q,T)=\alpha\,Q+\beta\,T$ and non-linear $f(Q,T)=Q+\lambda\,Q^2+\eta\,T$, in our study. Our aim is to find the exact wormhole solutions for both the models. Since our obtained field equations are more complex than solutions in classical GR, finding exact solutions with linear EoS and anisotropic relation is a challenging task for both the models. We have  found the exact solutions for both the cases under the assumption of linear $f(Q,T)$. For the case of non-linear model, finding the exact solutions is not analytically feasible for both equations of state. The solutions with linear EoS for linear form of $f(Q,T)$ are obtained explicitly. We have found that the shape function is in power-law form. Moreover, to satisfy the asymptotic flatness condition, some domains of $\omega$ and $\beta$ are found, which has shown in Table-\ref{table:1}. We picked up some particular values of $\omega$ and $\beta$ from table-\ref{table:1} and showed the behaviors of shape functions. We observed that all the necessary conditions of shape functions are satisfied, which is necessary for a traversable wormhole. Further, we checked the behavior of energy density in both phantom and quintessence regions and observed that it shows positively decreasing behavior in the phantom region, whereas it is violating in the quintessence region. Keeping this in mind, we picked up some particular values of $\omega$ and plotted the graphs for energy conditions. We observe that NEC is violated for radial pressure and is satisfied for tangential pressure. DEC is satisfied while SEC is violated throughout the spacetime. In the table-\ref{table:2}, we have summarised the behavior of energy conditions.\\
Furthermore, we have studied wormhole solutions for the anisotropic case for both models of $f(Q,T)$. Our analysis shows that for the non-linear model, it is tough to find exact solutions. For the linear model though, wormhole solutions for an anisotropic case are obtained and we obtain the shape function in power-law form. In table-\ref{table:3}, we have shown all the possible domains of $n$ and $\beta$ that satisfy the asymptotic flatness conditions. Figs. \ref{Fig6}-\ref{fig9} show that all the necessary conditions of shape functions are satisfied within each of the domains from the table. Moreover, the plot for energy density has been depicted in Fig. \ref{fig10} and one can observe that for some particular values of $n$, energy density is positive. Considering these values, we have plotted the graphs of NEC, DEC, and SEC. It is observed that NEC and SEC are violated, whereas DEC is obeyed throughout the spacetime in this $f(Q,T)$ gravity. The summary of the energy conditions for this case is depicted in table-\ref{table:4}.\\
As a matter of completeness, we checked the stability of our obtained wormhole solutions for both cases. From Figs. \ref{fig14}-\ref{fig15}, we can conclude that the wormhole solutions obtained by us are stable. In our whole study, we considered the redshift function $\phi(r)=constant$ to avoid the presence of an event horizon. It would be interesting to study wormhole solutions with non-constant redshift function and different matter sources  taken into account in this $f(Q,T)$ gravity.

\acknowledgments  M.T. acknowledges University Grants Commission (UGC), New Delhi, India, for awarding National Fellowship for Scheduled Caste Students (UGC-Ref. No.: 201610123801). Z.H. acknowledges the Department of Science and Technology (DST), Government of India, New Delhi, for awarding a Senior Research Fellowship (File No. DST/INSPIRE Fellowship/2019/IF190911). PKS acknowledges National Board for Higher Mathematics (NBHM) under Department of Atomic Energy (DAE), Govt. of India for financial support to carry out the Research project No.: 02011/3/2022 NBHM(R.P.)/R\&D II/2152 Dt.14.02.2022. We are very much grateful to the honorable referee and to the editor for the illuminating suggestions that have significantly improved our work in terms of research quality, and presentation.  


\end{document}